\documentclass[a4paper,twocolumn,10pt,accepted=2021-09-06]{quantumarticle}
\pdfoutput=1

\usepackage[american]{babel}
\usepackage{amssymb}
\usepackage{amsmath}
\usepackage{amsthm}
\usepackage{braket}
\usepackage[utf8]{inputenc}
\usepackage{graphicx}
\usepackage{caption}
\usepackage{csquotes}
\usepackage[allcolors=quantumviolet]{hyperref}
\usepackage{enumerate}
\usepackage{tikz}
\usetikzlibrary{quantikz}
\usepackage[normalem]{ulem}
\usepackage[backend=bibtex,style=phys,biblabel=brackets]{biblatex}
\AtEveryBibitem{\clearfield{note}}

\newcommand{\KL}{\ensuremath{\mathrm{KL}}}

\newcommand{\intstrut}{\vphantom{\int}}

\newif\ifverbose
\verbosetrue

\makeatletter
\renewcommand*\paragraph{\@startsection{paragraph}{4}{\z@}%
	{1.5ex \@plus1ex \@minus.2ex}
	{-1em}%
	{\normalfont\normalsize\itshape}}
\makeatother

\newcommand{\transp}{\ensuremath{\scriptscriptstyle T}}

\newcommand*{\Tr}{\operatorname{Tr}}

\providecommand{\myvec}[1]{\ensuremath{\boldsymbol{#1}}}

\providecommand{\ee}{\ensuremath{\myvec{e}}}
\providecommand{\ff}{\ensuremath{\myvec{f}}}

\providecommand{\tt}{\ensuremath{\myvec{t}}}

\providecommand{\vv}{\ensuremath{\myvec{v}}}

\providecommand{\ddelta}{\ensuremath{\myvec{\delta}}}

\providecommand{\ttheta}{\ensuremath{\myvec{\theta}}}

\providecommand{\mmu}{\ensuremath{\myvec{\mu}}}

\providecommand{\xxi}{\ensuremath{\myvec{\xi}}}

\providecommand{\pphi}{\ensuremath{\myvec{\phi}}}
\providecommand{\vvarphi}{\ensuremath{\myvec{\varphi}}}

\providecommand{\calE}{\ensuremath{\mathcal{E}}}
\providecommand{\calF}{\ensuremath{\mathcal{F}}}

\providecommand{\calM}{\ensuremath{\mathcal{M}}}

\providecommand{\bbE}{\ensuremath{\mathbb{E}}}

\providecommand{\bbI}{\ensuremath{\mathbb{I}}}

\providecommand{\bbR}{\ensuremath{\mathbb{R}}}

\begin{document}
\title{Fisher Information in Noisy Intermediate-Scale Quantum Applications}
\author{Johannes Jakob Meyer}
\affiliation{Dahlem Center for Complex Quantum Systems, Freie Universit\"{a}t Berlin, 14195 Berlin, Germany}
\affiliation{QMATH, Department of Mathematical Sciences, K{\o}benhavns Universitet, 2100 K{\o}benhavn {\O}, Denmark}
\orcid{0000-0003-1533-8015}
\date{26-06-2021}

\begin{abstract}
    The recent advent of noisy intermediate-scale quantum devices, especially near-term quantum computers, has sparked extensive research efforts concerned with their possible applications. At the forefront of the considered approaches are variational methods that use parametrized quantum circuits.
    The classical and quantum Fisher information are firmly rooted in the field of quantum sensing and have proven to be versatile tools to study such parametrized quantum systems.
    Their utility in the study of other applications of noisy intermediate-scale quantum devices, however, has only been discovered recently.
    Hoping to stimulate more such applications, this article aims to further popularize classical and quantum Fisher information as useful tools for near-term applications beyond quantum sensing. 
    We start with a tutorial that builds an intuitive understanding of classical and quantum Fisher information and outlines how both quantities can be calculated on near-term devices. We also elucidate their relationship and how they are influenced by noise processes. Next, we give an overview of the core results of the quantum sensing literature and proceed to a comprehensive review of recent applications in variational quantum algorithms and quantum machine learning.
\end{abstract}

\maketitle

\vspace{1em}

Progress in science and engineering as well as considerable investments have increased our capabilities to precisely control quantum systems, leading to the development of quantum devices that are capable of impressive feats. 
One prominent example of this new generation of devices are quantum computers. They have low numbers of qubits and are still plagued by noise and relatively low coherence times and cannot yet fulfill the promises of fault-tolerant quantum computation, but it was recently shown that they can outperform classical computers~\cite{arute2019quantum} -- however only in certain contrived tasks with unknown practical relevance. 
Yet, given these positive results and the speed of improvement of these devices, it is no surprise that the search for practically relevant applications of these \emph{noisy intermediate-scale quantum (NISQ)}~\cite{preskill2018quantum} computers has become a very active field of research in quantum information science. 
Particularly prominent candidates to make use of near-term quantum computers are \emph{variational quantum algorithms}~\cite{cerezo2020variational,bharti2021noisy} where parametrized quantum states are combined with a classical computer into a hybrid quantum-classical algorithm~\cite{mcclean2016theory}.
Another much researched direction is to use these quantum devices as \emph{quantum machine learning models}~\cite{benedetti2019parameterized}, again employing parametrized quantum states.

Along with the development of these new techniques, there is a need for a better \emph{understanding} of parametrized quantum systems and consequently the approaches that employ them. In the field of quantum sensing, a particular type of parametrized quantum system -- namely parametrized by the parameters that are sensed -- has long been studied. The principal tools employed in this regard are the \emph{classical} and \emph{quantum Fisher information}.

Intuitively, the quantum Fisher information is a measure of how much a parametrized quantum state changes under a change of a parameter. This parameter could be the underlying magnetic field that a quantum system is designed to sense, but it can also be a control knob turned by an experimenter or the angle of a rotation gate that is changed by a quantum computer programmer. The classical Fisher information in turn captures how much a change of the underlying parameter affects the probabilities with which different outcomes of a specific measurement that is performed on the parametrized quantum state are observed.

As parametrized quantum states appear in many approaches in NISQ applications beyond quantum sensing, it is no surprise that the first works have explored the use of classical and quantum Fisher information to study them. 
We do, however, believe that there are many more fruitful applications waiting to be discovered. This article thus intends to further popularize classical and quantum Fisher information as versatile tools to understand NISQ applications.

While much has been written about the Fisher information, in classical and quantum settings alike, the barriers to penetrate the literature can be quite substantial. This article aims to complement the excellent reviews provided in Refs.~\cite{liu2020quantum,sidhu2020geometric,katariya2021geometric} with a tutorial that provides a very gentle and intuitive introduction to both the classical and quantum Fisher information that does require only little prerequisites.
Since many of the results related to these quantities were developed in the context of quantum sensing, we will also review the results from this field that are most relevant to understanding applications of Fisher information. However, we will not explore specific applications, as it is not the goal of this article to provide a comprehensive introduction to quantum sensing.
We conclude this work with a comprehensive review of the different uses of Fisher information that have emerged in quantum machine learning and variational quantum algorithms.

The principal target audience of this paper are people that are interested in learning about Fisher information and its use in the context of NISQ applications. It thus only requires familiarity with the absolute basics of NISQ applications. The intuitive approach to the subject should also be beneficial to a wider class of readers, \emph{e.g.}\ people that are taking their first steps in quantum sensing.
After reading this paper, the reader will have an intuitive understanding of the origins and applications of both classical and quantum Fisher information and will know how they are related to each other. We will discuss how these quantities can be calculated in the context of NISQ applications, how the ever-present device noise enters the picture and how they are applied in quantum machine learning and optimization of variational quantum algorithms. Along the way, we will review important results from quantum sensing and try to demystify technical terms usually encountered when reading about Fisher information like \enquote{metric} or \enquote{pullback} and quantum-specific jargon like \enquote{Heisenberg scaling}.

This work is organized as follows: In Sec.~\ref{sec:preliminaries}, we will build intuition about parametrized quantum states and how we can properly assign distances to pairs of parameters. Sec.~\ref{sec:information_matrices} outlines how we can extract information about parametrized quantum systems in the form of information matrices. We follow up on this with the introduction of the classical Fisher information matrix in Sec.~\ref{sec:cfim} and the quantum Fisher information matrix in Sec.~\ref{sec:qfim}. We will put a particular focus on how we can actually compute these quantities in a NISQ context. The subsequent Sec.~\ref{sec:relation_cfim_qfim} elucidates the relationship between the quantum and the classical Fisher information, whereas Sec.~\ref{sec:noise} treats the role of noise, a very important impediment and namesake of NISQ devices. In Sec.~\ref{sec:nisq_applications} we will highlight the different areas in which the classical and quantum Fisher information have been applied in the context of NISQ devices, rounding up with a short outlook on fascinating applications beyond the scope of NISQ devices in Sec.~\ref{sec:beyond_nisq}. The work concludes with a look to the future in Sec.~\ref{sec:outlook}. To make this work as self-contained and explanatory as possible, most derivations are found in the Appendix, alongside with proofs of the properties of classical and quantum Fisher information.
In the course of this work we will give recommendations on literature for further study.

\section{Parametrized Quantum States}\label{sec:preliminaries}
The study of NISQ devices places a huge emphasis on the study of \emph{parametrized quantum states}, \emph{i.e.}\ quantum states that depend continuously on a vector of parameters $\ttheta \in \bbR^d$. We will denote them with $\ket{\psi(\ttheta)}$ for pure states and $\rho(\ttheta)$ for mixed states. 
Parametrized quantum states arise in many settings: In NISQ computing, \emph{parametrized quantum circuits} are used as an \emph{ansatz} for variational quantum algorithms~\cite{cerezo2020variational}. In quantum optimal control, the parameters of the radio frequency pulses determine the executed gate or the created quantum state. In quantum metrology, a quantity that needs to be measured, for example a magnetic field, is imprinted on a quantum state via an interaction, thus parametrizing it.
 
It is of course of tremendous interest to understand what happens if the parameters $\ttheta$ are changed. We can attempt to quantify this by measuring the \emph{distance} between the parameters themselves -- which can be done via the regular Euclidean distance. But only in the rarest cases all parameters have an equal influence on the underlying state. So a smarter way to measure distances between parameters would actually be to measure the distance of the associated \emph{states}. If we are given a distance measure $d$ between quantum states, we can define -- by a slight abuse of notation -- a new distance measure between the associated parameters:
\begin{align}
    d(\ttheta, \ttheta') = d(\rho(\ttheta), \rho(\ttheta')).
\end{align}
This strategy of measuring the distance in the space of quantum states instead of the parameter space is known as a \emph{pullback}, because we pull back the distance measure to the space of parameters.

In the following we will be concerned with distance measures that are \emph{monotonic}. This means that the distance between two quantum states can only decrease if both are subject to the same quantum operation. A quantum operation can take many forms, including unitary evolution, noisy evolution or even measurements -- but we will formalize this more later on.  
Monotonicity is a very desirable property for a distance measure. Performing a quantum operation cannot add additional information, so it should not be easier to distinguish two states after such an operation is performed. Moreover, if the operation is noisy, it will actually \emph{destroy} information, which should be echoed in a decrease of the associated distance as adding noise cannot make two states more distinguishable~\cite{nielsen2010quantum}.

\begin{figure}
    \centering
    \includegraphics[width=\linewidth]{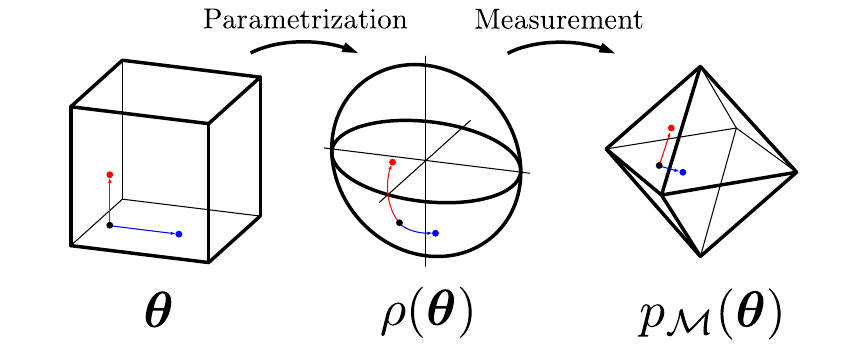}
    \caption{Large distances between parameters $\ttheta$ and $\ttheta'$ need not correspond to large distances of the corresponding quantum states $\rho(\ttheta)$ and $\rho(\ttheta')$. Equally, large distances between quantum states need not correspond to large distances between the output probability distributions after the measurement, $p_{\calM}(\ttheta)$ and $p_{\calM}(\ttheta')$. Measuring the distance between two parameters by measuring the distance between the corresponding quantum states or output probability distributions is therefore a sensible approach known as a \emph{pullback}.}
    \label{fig:pullback_distance}
\end{figure}
We have already argued that it is more sensible to measure distances between parameters of a quantum state by measuring the distances in the space of quantum states. But as we are mere classical observers, any NISQ application must involve a measurement of the underlying quantum state. A measurement will inevitably collapse the quantum state into a classical probability distribution over the possible measurement outcomes. We can formalize this by defining a measurement $\calM = \{ \Pi_l \}$, where the operator $\Pi_l$ identifies the $l$-th outcome of the experiment~\cite{nielsen2010quantum}. In NISQ applications, the operators $\{ \Pi_l \}$ are usually just the projectors onto the basis states. The probabilities of observing the different outcomes are then given by
\begin{align}
    p_l(\ttheta) = \Tr \{ \rho(\ttheta) \Pi_l \}.
\end{align}
We will use $p_{\calM}(\ttheta)$ to refer to the full output distribution for a specific measurement $\calM$. We will drop the subscript $\calM$ if we talk about generic probability distributions.

If we fix a certain measurement $\calM$, we also fix a certain way of collapsing a parametrized quantum state into a probability distribution which is now also dependent on the parameter $\ttheta$. We thus have a \emph{parametrized probability distribution}. But now the same argument that we used to motivate measuring the distance between parameters in the space of quantum states applies here, too. A large change in the underlying quantum state might not correspond to a large change in the output probability distribution that is observed. If we fix a measurement, we should therefore consider the pullback of a distance $d$ between two probability distributions
\begin{align}
    d_{\calM}(\ttheta, \ttheta') = d(p_{\calM}(\ttheta), p_{\calM}(\ttheta')).
\end{align}

In the following, we require some very natural properties from the distance measures between quantum states or probability distributions we employ. First, that it is always positive $d(\ttheta, \ttheta') \geq 0$ and second, that the distance between identical objects is zero $d(\ttheta, \ttheta) = 0$. This means that the distance measure has to satisfy the axiomatic definition of a \emph{divergence} as employed in statistics, which is less strict than the definition of distance encountered in other areas of mathematics.

In summary, we have now set the arena: the parameters $\ttheta$ define a quantum state $\rho(\ttheta)$ which then undergoes the measurement $\calM$. We can measure distances between parameters by going to the space of quantum states or to the space of probability distributions over the measurement outcomes, dependent on the question we seek to answer. Fig.~\ref{fig:pullback_distance} illustrates these relations.

\section{Information Matrices}\label{sec:information_matrices}
We are often confronted with scenarios where a quantum system is in a state associated to a particular parameter $\ttheta$ but where it is important to understand how much a change of the parameter $\ttheta$ in a particular direction results in a change of the underlying quantum state or the output probability distribution. 

To gain that understanding, we will look how a slight perturbation of the parameter $\ttheta + \ddelta$ reflects in the chosen distance by analyzing $d(\ttheta, \ttheta + \ddelta)$. If the distance measure $d$ is \emph{differentiable}, we can develop this into a Taylor series around $\ddelta = 0$. Because $d$ is a distance measure, we can assume that it is both positive and vanishes for identical parameters, \emph{i.e.}\ $d(\ttheta, \ttheta) = 0$ is a minimum. But we know that the first order contributions vanish around minima and that the second order is therefore the first contribution of the Taylor series that does not vanish. To write down the Taylor expansion, we first define the matrix $M$ with entries
\begin{align}\label{eqn:def_pullback_metric}
    M(\ttheta)_{ij} = \left. \frac{\partial^2 }{\partial \delta_i \partial \delta_j}d(\ttheta, \ttheta + \ddelta)\right|_{\ddelta = 0}.
\end{align}
In more mathematical terms, this is the matrix of second order derivatives -- also known as the \emph{Hessian} -- of the function 
\begin{align}
    g_{\ttheta}(\ddelta) = d(\ttheta, \ttheta + \ddelta),
\end{align}
at $\ddelta = 0$. With this we can express the Taylor expansion as
\begin{align}
    d(\ttheta, \ttheta + \ddelta) &= \frac{1}{2}\sum_{i,j=1}^d \delta_i \delta_j M(\ttheta)_{ij} + O(\lVert \ddelta \rVert^3) \\ 
    &= \frac{1}{2}\ddelta^{\transp} M(\ttheta) \ddelta + O(\lVert \ddelta \rVert^3). \label{eqn:distance_second_order}
\end{align}
The matrix $M(\ttheta)$ therefore captures all we need to know about the local vicinity of $\ttheta$ in parameter space, but measured by the distance $d$ in the underlying space of either quantum states or output probability distributions! Intuitively, large entries of $M(\ttheta)$ indicate that a change in the corresponding parameters results in a large change of the underlying object.
In the following, we will not denote the dependence on $\ttheta$ explicitly and simply use the notation $M$.

The matrix $M$ is an example of a \emph{metric}. To elucidate the meaning of this, we first need to remind ourselves how measuring distances and angles in Euclidean space works through the standard scalar product. It is defined as
\begin{align}
    \langle \ddelta, \ddelta' \rangle = \ddelta^{\transp} \ddelta'= \sum_{i=1}^d \delta_i^{\vphantom{'}} \delta_i'.
\end{align}
The standard scalar product acts as an \enquote{umbrella} of sorts as it allows us to measure lengths
\begin{align}
    \lVert \ddelta \rVert = \sqrt{\langle \ddelta, \ddelta \rangle},
\end{align}
distances
\begin{align}
    d(\ddelta, \ddelta') = \sqrt{\langle \ddelta - \ddelta', \ddelta - \ddelta' \rangle},
\end{align}
and angles between vectors
\begin{align}
    \sphericalangle(\ddelta, \ddelta') = \arccos \frac{\langle \ddelta, \ddelta' \rangle}{\sqrt{\langle \ddelta, \ddelta \rangle \langle \ddelta', \ddelta' \rangle}}.
\end{align}

It is quite suggestive that $\ddelta^{\transp} \ddelta' = \ddelta^{\transp} \bbI \ddelta'$ resembles the expression in Eq.~\eqref{eqn:distance_second_order} with $M$ replaced by the identity matrix $\bbI$. And indeed, we can use the matrix $M$ to define a new scalar product
\begin{align}
    \langle \ddelta, \ddelta' \rangle_M = \ddelta^{\transp} M \ddelta'
\end{align}
that now includes information about the local environment of $\rho(\ttheta)$ or $p_{\calM}(\ttheta)$! And this is what a metric does in a nutshell -- it allows to measure distances and angles between vectors in parameter space, but with the local structure of the underlying quantum states or probability distribution parametrized by those vectors taken into account.

Because these matrices contain information about the underlying quantum states and probability distributions -- which are by nature objects that have an information theoretic meaning -- we will call them \emph{information matrices}.

\section{The Classical Fisher Information}\label{sec:cfim}
After having kept the derivations general, we now turn our attention to the Fisher information itself and begin with the classical case.
As the name suggests, the classical Fisher information is defined for parametrized probability distributions. In the context of NISQ devices, this means on the probability distributions of measurement outcomes. To make use of the machinery we developed in the last section, we need a distance measure between probability distributions. There exists numerous ways to measure distances between probability distributions, but one of the most popular is certainly the \emph{Kullback-Leibler (KL) divergence}, also known as the \emph{relative entropy}. It is defined as\footnote{Note that there are edge cases where the KL divergence is not properly defined, for example if there exists a $p_l(\ttheta') = 0$ where at the same time $p_l(\ttheta) \neq 0$. We will exclude this cases in this work.}
\begin{align}\label{eqn:def_kl_divergence}
    d_{\KL}(p_{\calM}(\ttheta), p_{\calM}(\ttheta')) = \sum_{l \in \calM} p_l( \ttheta) \log \frac{p_l(\ttheta)}{p_l(\ttheta')}.
\end{align}
The intuition behind the KL divergence is not obvious at first sight, but Nielsen and Chuang give some accounts in Ref.~\cite{nielsen2010quantum}.
Imagine we are given an unknown probability distribution that can be either $p_{\calM}(\ttheta)$ or $p_{\calM}(\ttheta')$ and we are tasked with deciding which of the two distributions it is. Then, the KL divergence essentially captures how fast the \enquote{false negative} error decreases with the number of repetitions we are allowed when there is a constraint on the \enquote{false positive} probability.

Now, we need to perform the second order expansion to get a formula for the corresponding information matrix. For conciseness of the main text, you find the complete derivation in App.~\ref{app:deriv_cfim}. The result of the calculation is the following formula for the information matrix associated to the KL divergence:
\begin{align}\label{eqn:classical_fisher_formula}
    [M_{\KL}]_{ij} &=  \sum_{l \in \calM }p_l(\ttheta)   \frac{\partial^2 }{\partial \theta_i \partial \theta_j}\log p_l(\ttheta) \\
    &= \sum_{l \in \calM } \frac{1}{p_l(\ttheta)} \frac{\partial p_l(\ttheta)}{\partial \theta_i}\frac{\partial p_l(\ttheta)}{\partial \theta_j}.
\end{align}
The matrix $M_{\KL}$ is nothing else than the \emph{(classical) Fisher information matrix (CFIM)} and we will in the following denote it as $I = M_{\KL}$. A resource with substantial information about it is the textbook by Lehmann and Casella~\cite{lehmann1998theory}. For the reader's convenience, we list and prove the most important properties of the classical Fisher information matrix in App.~\ref{app:properties_cfim}.

\paragraph{Uniqueness.}
We derived the classical Fisher information from the Kullback-Leibler divergence. But what happens if we repeat the same procedure for another distance measure? It turns out that the derivation will \emph{always} yield a constant multiple of the classical Fisher information if the distance measure is monotonic. 

We shallowly introduced the idea of monotonicity in Sec.~\ref{sec:preliminaries}, but we will formalize it now. To this end, we first need another concept, namely that of a \emph{stochastic map}. A stochastic map is a linear operation that takes in probability distributions and always outputs probability distributions. This is a very broad definition and includes every admissible thing we can do with probability distributions. We formally call a distance measure between two probability distributions $p$ and $q$ monotonic if the distance cannot increase under any stochastic map $T$:
\begin{align}
    d(T[p], T[q]) \leq d(p, q).
\end{align}

The fact that we always end up with a constant multiple of the Fisher information if we start from a monotonic distance measure is known as the \emph{uniqueness} of the classical Fisher information which is a celebrated result due to Morozova and Chentsov~\cite{morozova1991markov}. 
A particularly important and widely applied monotonic distance measure between probability distributions is given by the \emph{total variation distance}
\begin{align}
    d_{\mathrm{TV}}(p, q) = \frac{1}{2} \sum_{l} |p_l - q_l|.
\end{align}
It measures the maximum difference in probability that $p$ and $q$ can assign to the same event. It would be tempting to ask which information matrix is related to this distance measure. But we have to remember that we required the distance measure to be differentiable in order to define an information matrix -- and as the total variation distance includes an absolute value function, it is not differentiable at $p = q$ as would be required for our construction. The total variation distance therefore does not induce an information matrix.

\paragraph{Calculation.}
To work with the classical Fisher information matrix in a NISQ context we need to actually calculate it. If we take a closer look at Eq.~\eqref{eqn:classical_fisher_formula}, we see that we need two ingredients to do so: First, the probabilities of the different measurement outcomes $p_l(\ttheta)$ and second their derivatives with respect to the parameters $\ttheta$, $\partial p_l(\ttheta)/\partial \theta_i$.

First, let us talk about the output probabilities.
Each run of a NISQ device with a fixed measurement setting will add a data point, and with sufficiently many data points, the output probabilities $p_l(\ttheta)$ can be estimated. This is also how one approaches the estimation of expectation values: a unitary transformation changes the measurement basis to the eigenbasis of the desired operator $H = \sum_l h_l |h_l \rangle \! \langle h_l |$. We then perform multiple repetitions of the experiment, estimate the output probabilities $p_l(\ttheta)$ and then calculate the expectation value as
\begin{align}
    \langle H(\ttheta) \rangle = \sum_l p_l(\ttheta) h_l.
\end{align}
This means that an estimation of the output probabilities is actually a prerequisite for the calculation of expectation values, the most ubiquitous primitive in NISQ settings. Note however that the output probability distribution contains more information than the expectation value -- which means that a faithful estimate of the probability distribution will usually require more runs of the experiment than an estimation of an expectation value. 

In fact, in the worst case the number of samples required to estimate the probability distribution is proportional to the number of different measurement outcomes which usually is exponential in the number of qubits~\cite{canonne2020short}. This problem is ameliorated if many of the output probabilities are very small -- in this case we need fewer samples to guarantee a good estimate. This can be understood intuitively: If we do not observe a particular measurement outcome $l$, we estimate $p_l = 0$. But as we know that $p_l$ is small, this is already a good estimate.

As the estimation of the output probability distribution is a very important task in NISQ settings, there also exist more sophisticated methods than just estimating the output probabilities from the number of times they were observed in a limited number of test runs. One possibility is a Bayesian approach, where a prior estimate of the output distribution is updated using new samples. Techniques based on machine learning have also been shown to be very effective tools to capture information about quantum states and therefore also the associated output probability distributions~\cite{torlai2020machine-learning}.

We now turn to the derivatives. In many cases, gradient based schemes are deployed to optimize Variational Quantum Algorithms. We will now argue that these schemes already entail the calculation of the derivatives of the output probabilities necessary for the calculation of the Fisher information matrix. These methods use either finite differences or the \emph{parameter-shift rule}~\cite{mitarai2018quantum,schuld2019evaluating}. In all cases, the derivatives of expectation values are calculated by evaluating expectation values at different parameter settings. We will make this exemplary by considering two-sided finite differences, where the derivative is approximated via a small perturbation $\epsilon$:
\begin{align}
    \frac{\partial}{\partial \theta_i} \langle H(\ttheta) \rangle \approx \frac{\langle H(\ttheta + \epsilon \ee_i)-\langle H(\ttheta - \epsilon \ee_i) \rangle}{2 \epsilon},
\end{align}
where $\ee_i$ denotes the $i$-th unit vector, or, equivalently, that we only perturb $\theta_i$. The following argument, however, equally works for other means of estimating derivatives.

In order to perform the finite-difference approximation, we have to compute the expectation values $\langle H(\ttheta \pm \ee_i \epsilon)\rangle$, which proceeds via the estimation of the output probability distribution as explained above. But this means that we actually compute the derivative of the output probability distribution:
\begin{align}
    \frac{\partial}{\partial \theta_i} \langle H(\ttheta) \rangle &\approx \frac{\sum_l p_l(\ttheta + \epsilon \ee_i) h_l-\sum_l p_l(\ttheta - \epsilon \ee_i) h_l}{2 \epsilon} \\
    &= \sum_l \frac{p_l(\ttheta + \epsilon \ee_i) - p_l(\ttheta - \epsilon \ee_i)}{2 \epsilon} h_l \\
    &\approx \sum_l \frac{\partial p_l(\ttheta)}{\partial \theta_i} h_l.
\end{align}
This means that the same process we use to estimate derivatives of expectation values will also yield the derivatives of the output probabilities we need to calculate the classical Fisher information matrix. Again a word of caution is advised, as the faithful estimation of the derivatives of the whole output probability distribution will usually require more runs of the experiment than used for estimating the derivative of an expectation value. 

In summary, we learned that the processes that are already used to compute expectation values and their gradients implicitly give us the data necessary to estimate both the output probabilities $p_l(\ttheta)$ and their derivatives $\partial p_l(\ttheta)/\partial \theta_i$ and thus the whole classical Fisher information matrix. This brings us to the conclusion that the calculation of the Fisher information matrix is not harder than the calculation of expectation values and their derivatives which are routine tasks in the context of NISQ computing.

We should also note that -- especially for large quantum systems -- one usually does not have sufficiently many samples to accurately estimate the whole distribution, which means that many elements of the probability distribution $p_l$ are estimated as zero. In this case, the corresponding terms in the formula for the classical Fisher information of Eq.~\eqref{eqn:classical_fisher_formula} would diverge. This is usually mitigated by the fact that the associated derivatives are also zero in which case we can just remove the corresponding terms from the sum. At points where $p_l$ vanishes but a derivative $\partial p_l(\ttheta)/\partial \theta_i$ does not we actually have a discontinuity of the Fisher information~\cite{seveso2020discontinuity}, which means that it is not properly defined at these points, a property that is inherited from the Kullback-Leibler divergence. Another way around this problem is to use a Bayesian approach for the estimation of the output probability distribution with a prior that does not contain zero probabilities.

\section{The Quantum Fisher Information}\label{sec:qfim}
We always strive to find quantum generalizations of classical concepts, and the Fisher information is no exception. We know that any classical probability distribution can be expressed by some quantum state with a diagonal density matrix, which means that classical probability distributions are actually a subset of all quantum states. Because of the uniqueness of the classical Fisher information, any \enquote{quantum Fisher information} should reduce to the classical Fisher information when looking on classical states. 

We have furthermore learned that the classical Fisher information is associated with \emph{monotonic} distance measures. To properly define monotonicity in the quantum setting we first need to introduce the quantum generalization of stochastic maps, the \emph{quantum channels}. Stochastic maps are linear operations that map probability distributions to probability distributions. Likewise, quantum channels are defined as linear operations that take density matrices in and output density matrices, even when ancillary systems are included. Like stochastic maps, quantum channels are a very broad concept, including not only unitary and noisy evolution but also measurements! This can actually be seen quite easily: measurements turn a quantum state into a classical probability distribution over the measurement outcomes. But as we just learned these are also a subset of all quantum states, which means that measurements match the requirements of a quantum channel. 
A distance measure $d$ between quantum states is monotonic if it cannot increase under any quantum channel $\Phi$:
\begin{align}
    d(\Phi[\rho], \Phi[\sigma]) \leq d(\rho, \sigma).
\end{align}
As quantum channels encompass basically everything we can do in quantum information processing, they are heavily studied. You find introductions to quantum channels in the excellent book by Wilde~\cite{wilde2017classical}, the classic book by Nielsen and Chuang~\cite{nielsen2010quantum} and the more mathematical lecture notes by Wolf~\cite{wolf2014quantum}.

To find a quantum generalization of the classical Fisher information, we will again use the machinery developed in Sec.~\ref{sec:information_matrices}. We will limit ourselves to \emph{pure} quantum states in this section to keep the developments simple, but return to the general case in Sec.~\ref{sec:noise}. 

To start the machinery, we need to choose an appropriate distance measure between quantum states. From all possible contenders, the \emph{fidelity} stands out due to its beautiful operational interpretation. The fidelity between two pure quantum states is given by
\begin{align}
    f(\ket{\psi(\ttheta)}, \ket{\psi(\ttheta')}) = | \langle \psi(\ttheta) | \psi(\ttheta') \rangle |^2.
\end{align}
The fidelity is so important because the probability with which we can distinguish the two states $\ket{\psi(\ttheta)}$ and $\ket{\psi(\ttheta')}$ when using the optimal measurement is 
\begin{align}
    d_f(\ket{\psi(\ttheta)}, \ket{\psi(\ttheta')}) = 1 - f(\ket{\psi(\ttheta)}, \ket{\psi(\ttheta')}).
\end{align}
As we need a measure of distance that is 0 for indistinguishable states, we will use $d_f$ in our calculations.\footnote{We could also have used a different convention for our calculations, where the square root of the fidelity is used instead of the fidelity. App.~\ref{app:rescaling} contains an explanation why this will only incur a constant prefactor.}

We will again leave the complete derivation to App.~\ref{app:deriv_qfim} and directly go to the formula for the information matrix associated with the fidelity:
\begin{equation}\label{eqn:information_matrix_fidelity}
\begin{split}
    [M_f]_{ij} &= 2 \operatorname{Re}\big[\langle\partial_i \psi(\ttheta) | \partial_j \psi(\ttheta) \rangle \\
    &\qquad- \langle\partial_i \psi(\ttheta) | \psi(\ttheta) \rangle \langle \psi(\ttheta) | \partial_j \psi(\ttheta) \rangle\big].
\end{split}
\end{equation}
In this formula, we have used the shorthand $\partial_i = \partial/\partial \theta_i$.
If one checks the consistency of this information matrix with the classical Fisher information, however one realizes that the prefactor is wrong. As this is only an artifact of how defined our distance we can simply correct this by multiplying with a constant. Doing so, we obtain the formula for the \emph{quantum Fisher information matrix (QFIM)} which is associated with the distance $2 d_f$:
\begin{equation}\label{eqn:def_qfim}
\begin{split}
    \calF_{ij} &= 4 \operatorname{Re}\big[\langle\partial_i \psi(\ttheta) | \partial_j \psi(\ttheta) \rangle \\
    &\qquad\qquad- \langle\partial_i \psi(\ttheta) | \psi(\ttheta) \rangle \langle \psi(\ttheta) | \partial_j \psi(\ttheta) \rangle\big].
\end{split}
\end{equation}

A great review about the quantum Fisher information matrix, various ways to calculate it and many applications was recently written by Liu \emph{et al.}~\cite{liu2020quantum}. For the reader's convenience, however, the important properties of the quantum Fisher information matrix are summarized and proven in App.~\ref{app:properties_qfim}.

\paragraph{Non-Uniqueness.}
In the classical case, it does not matter from which monotonic distance measure between probability distributions we start our derivation, we will always end up with a constant multiple of the classical Fisher information. But this is actually not the case in the quantum setting! Indeed, as was proven by Pétz~\cite{petz1996monotone}, there are infinitely many monotonic metrics in the quantum setting, and multiple ones have found application in quantum information theory. To distinguish the quantum Fisher information that arises from the fidelity from the other ones it is also often referred to as \emph{SLD quantum Fisher information}. Here, SLD stands for \emph{symmetric logarithmic derivative}. This is related to a different way in which we can define the (SLD) quantum Fisher information. It is 
\begin{align}
    \calF_{ij} = \frac{1}{2}\Tr \{ \rho (L_i L_j + L_j L_i) \}
\end{align}
where $L_i$ is the \emph{symmetric logarithmic derivative (SLD) operator} corresponding to the coordinate $\theta_i$. It is implicitly defined by
\begin{align}\label{eqn:sld_operator_implicit_def}
    \frac{\partial \rho}{\partial \theta_i} = \frac{1}{2}(L_i \rho + \rho L_i).
\end{align}
You can think of this operator as a way to rewrite derivatives of the quantum state $\rho$. We chose to not introduce the quantum Fisher information via this approach because it is not only unintuitive but also somewhat unwieldy. This, however, is the way in which the first results on the quantum Fisher information were obtained~\cite{helstrom1967minimum}. 

Pétz also showed that the (SLD) quantum Fisher information that we just derived from the fidelity between quantum states is special because it is the \emph{smallest} monotone metric in a certain sense~\cite{petz1996monotone}.
One can furthermore make a case that it is the \enquote{most natural} in certain respects. If your are interested in this, have a look at Ref.~\cite{cheng2013quantum} which contains a pedagogical exposition relating it to the \enquote{natural} geometry of the Hilbert space of quantum states. 
Another useful property of the SLD quantum Fisher information that sets it apart from the competition is that it is actually defined for pure states -- many other possible generalizations of the classical Fisher information become infinite in this case. 

\paragraph{Calculation.}
The quantum Fisher information is a much more peculiar object to work with than its classical counterpart. We will now outline the techniques that have been developed to tackle the calculation of the quantum Fisher information. We will keep our focus on the pure state case and come back to the practically important noisy case later. 

In many NISQ applications, especially in NISQ computing, the parameters of the quantum state are usually rotation angles of gates with a certain generator, \emph{e.g.}\ $U(\theta_i) = e^{-i \theta_i G_i}$. In this case, our job will be a lot easier. If we execute the circuit that prepares our state until the point where the gate in question is applied and call the state before the gate happens $\ket{\psi_0}$ we can express the derivative in terms of the generator:
\begin{align}
    |\partial_i \psi(\ttheta) \rangle = \partial_i e^{-i \theta_i G_i} \ket{\psi_0} = -i G_i \ket{\psi(\ttheta)}.
\end{align}
Putting this into the formula for the quantum Fisher information of Eq.~\eqref{eqn:def_qfim}, we get a simple formula for the diagonal elements:
\begin{align}
    \calF_{ii} = 4\left(\langle \psi_0 | G_i^2 | \psi_0 \rangle - \langle \psi_0 | G_i | \psi_0 \rangle^2\right).
\end{align}
This is nothing but the fourfold variance of the generator $G_i$ with respect to the state $\ket{\psi_0}$. Note that we dropped the real part from the formula as the variance is already a real number. This means that we can get the diagonal elements of the quantum Fisher information matrix by executing the circuit in question until our gate happens and then evaluating the expectation values of the two observables $G_i^2$ and $G_i$.

If we have multiple gates happening in parallel, we can also use the same approach to compute the off-diagonal elements of the quantum Fisher information matrix related to these gates. If we evaluate the formula for the quantum Fisher information in this case, we get
\begin{align}
\begin{split}
    \calF_{ij} = 4\Big(&\langle \psi_0 |\frac{\{G_i, G_j\}}{2} |\psi_0 \rangle \\
    &- \langle \psi_0 | G_i | \psi_0 \rangle\langle \psi_0 | G_j | \psi_0 \rangle \Big),
\end{split}
\end{align}
where $\{G_i, G_j\}/2 = (G_i G_j + G_j G_i)/2$ can be understood as the \enquote{real part} of the product $G_i G_j$. The quantity above is nothing else but the fourfold \emph{covariance} of the generators $G_i$ and $G_j$ with respect to the state $\ket{\psi_0}$. This means that we can evaluate all \enquote{blocks} of the quantum Fisher information matrix corresponding to gates executed in parallel by evaluating the aforementioned observables on the state right before the layer of parallel gates is executed, $\ket{\psi_0}$~\cite{stokes2020quantum}.

Elements of the quantum Fisher information matrix that correspond to gates that are not executed in parallel are harder to deal with, because the observables that need to be evaluated now also depend on the intermediary circuit elements between the gates. But we can still evaluate the elements of the quantum Fisher information in this case using more sophisticated techniques. 

If the quantum gates that shall be differentiated support a \emph{parameter-shift rule} we can do this across layers. A parameter-shift rule~\cite{schuld2019evaluating} states that the expectation value of any operator $O$ evaluated on a state $\ket{\psi(\ttheta)}$, $\langle O(\ttheta) \rangle = \langle \psi(\ttheta) | O | \psi(\ttheta) \rangle$ can be differentiated as
\begin{align}
    \frac{\partial}{\partial \theta_i}\langle O(\ttheta) \rangle = r\left(\langle O(\ttheta + \ee_i\frac{\pi}{4r}) \rangle - \langle O(\ttheta -\ee_i \frac{\pi}{4r}) \rangle \right)
\end{align}
where the constant $r$ depends on the nature of the gate in question. We see that the derivative can be evaluated by running the same circuit but \enquote{shifting} the parameter in question in both directions. Luckily, most quantum gates available on near-term quantum devices support such a parameter-shift rule. 

Using the parameter-shift rule and the fact that the quantum Fisher information matrix can be expressed via the second derivatives of the fidelity, the authors of Ref.~\cite{mari2021estimating} derived a formula for the quantum Fisher information that reads
\begin{equation}
\begin{split}
    \calF_{ij} = -\frac{1}{2}\Big(
    &|\langle \psi(\ttheta) | \psi(\ttheta + (\ee_i + \ee_j)\frac{\pi}{2}) \rangle|^2 \\
    -\mathstrut & |\langle \psi(\ttheta) | \psi(\ttheta + (\ee_i - \ee_j)\frac{\pi}{2}) \rangle|^2 \\
    -\mathstrut & |\langle \psi(\ttheta) | \psi(\ttheta - (\ee_i - \ee_j)\frac{\pi}{2}) \rangle|^2 \\
    +\mathstrut &|\langle \psi(\ttheta) | \psi(\ttheta - (\ee_i + \ee_j)\frac{\pi}{2}) \rangle|^2
    \Big).
\end{split}
\end{equation}
This formula contains the fidelities between different parametrizations of the same state. To evaluate those via quantum circuits, we have two principal ways: if we denote the circuit preparing the quantum state $\ket{\psi(\ttheta)}$ as $U(\ttheta)$, then the overlap $|\langle \psi(\ttheta) | \psi(\ttheta') \rangle|^2$ can be evaluated via first applying the unitary $U(\ttheta)$ and then the inverse of $U(\ttheta')$. After this circuit, the overlap is exactly the probability of observing the outcome 0~\cite{havlicek2018supervised}:
\begin{center}
    \begin{quantikz}
        \lstick{$\ket{0}$} & \gate{U(\ttheta)\intstrut}& \gate{U^{\dagger}(\ttheta')\intstrut} & \meterD{0\intstrut}
    \end{quantikz}    
\end{center}
Alternatively, we can make use of the \emph{SWAP test} which achieves the same thing:
\begin{center}
    \begin{quantikz}
        \lstick{$\ket{0}$} & \gate{H\intstrut}& \ctrl{2} & \gate{H\intstrut} & \meterD{\langle Z \rangle \intstrut} \\
        \lstick{$\ket{0}$} & \gate{U(\ttheta)\intstrut} & \swap{1} & \qw &  \\
        \lstick{$\ket{0}$} & \gate{U^{\dagger}(\ttheta')\intstrut} &\targX{} &  \qw &  \\
    \end{quantikz}    
\end{center}
Both approaches have their downsides. The compute and reverse approach has twice the depth of the original state preparation and the SWAP test requires twice as many qubits, an additional ancilla and controlled SWAP operations but retains the same depth up to some constant number of gates. We should also not forget that these approaches only work for pure states. As soon as the operations preparing the state become noisy, the quantities calculated via these approaches do not coincide with the fidelity but instead with the state overlap which does not give rise to the quantum Fisher information.

The property that the quantum Fisher information is associated to the second derivatives of the fidelity can also be used to get an approximation to the quantum Fisher information matrix via a finite differences approach. We have introduced the quantum Fisher information matrix as the second derivative of twice the fidelity distance. With finite differences, we can approximate the second derivative of any function $g(t)$ as 
\begin{align}
    \partial_{t}^2 g(t) \approx \frac{g(t + \epsilon) - 2 g(t) + g(t - \epsilon)}{\epsilon^2}.
\end{align}
If we now set
\begin{align}
    g(t) = 2 d_f(\ttheta, \ttheta + t \vv)
\end{align}
for some arbitrary unit vector $\vv$, we see that $g(0) = 0$ because $d_f$ is a distance and $g(t + \epsilon) = g(t-\epsilon)$ for small $\epsilon$, because the second order is the first non-vanishing order of the Taylor expansion of $d_f$.
Putting all of this together allows us to compute the projection of the quantum Fisher information matrix in a particular direction: 
\begin{align}
    \vv^{\transp} \calF \vv \approx \frac{4 d_f(\ttheta, \ttheta + \epsilon \vv)}{\epsilon^2},
\end{align}
where $\vv$ is an arbitrary vector of unit length and $\epsilon$ is small. We can therefore use quantum circuits that calculate the overlap between two pure states along with small perturbations to approximate the quantum Fisher information matrix. Note that you might also find other formulas that contain the square root of the fidelity. As argued in App.~\ref{app:rescaling}, these formulas are equally valid and stem from a different convention for the fidelity distance.

Recently, Ref.~\cite{gacon2021simultaneous} followed this spirit of approximation to generalize the \emph{simultaneous perturbation stochastic approximation (SPSA)} method for the stochastic approximation of gradients to the calculation of the quantum Fisher information matrix.
The original SPSA method computes estimates of the gradient of a function $f(\ttheta)$ using the finite differences approximation with small \emph{random} perturbations of the parameters. The same strategy can be applied to the finite differences expression for the second derivative, where two distinct random perturbations are used to obtain an estimate of the Hessian matrix. Applied to the estimation of the quantum Fisher information matrix, the technique proceeds by first selecting two random vectors of unit length, $\vv_1$ and $\vv_2$. Then, for a small $\epsilon$, the following quantity is computed:
\begin{align}
\begin{split}
    \delta \calF &= f_{\ttheta}(\epsilon \vv_1 + \epsilon \vv_2) -f_{\ttheta}(- \epsilon \vv_1) \\
    &-f_{\ttheta}( - \epsilon \vv_1 + \epsilon \vv_2) +f_{\ttheta}( + \epsilon \vv_1).
\end{split}
\end{align}
The shorthand $f_{\ttheta}(\ddelta)$ denotes the fidelity of the system state with the state whose parameters are perturbed by $\ddelta$,
\begin{align}
    f_{\ttheta}(\ddelta) = |\langle \psi(\ttheta) | \psi(\ttheta + \ddelta) \rangle|^2.
\end{align}
From this quantity, we can use the outer products of the perturbation vectors to generate an approximation of the quantum Fisher information matrix as
\begin{align}
    \hat{\calF} = -\frac{\delta \calF}{2 \epsilon^2}(\vv_1 \vv_2^{\transp} + \vv_2 \vv_1^{\transp}).
\end{align}
Note that the authors of Ref.~\cite{gacon2021simultaneous} use a different convention for the quantum Fisher information matrix that causes a difference in the prefactor. This approximation is not enough to faithfully estimate the whole quantum Fisher information matrix because it has a maximal rank of 2. But we can average multiple such estimates to get an approximation of the quantum Fisher information matrix with increasing quality.

\section{Relation of Classical and Quantum Fisher Information}\label{sec:relation_cfim_qfim}
To clarify the relation between classical and quantum Fisher information, we first return to the notion of \emph{monotonicity} that we required for the distance measures we look at. Intuitively, we would expect that the associated information measure also \enquote{decreases} if a quantum channel is applied to the underlying quantum system. 

And indeed, the monotonicity of the distance measure carries over to the associated information matrix. To elucidate \emph{how} this happens, we will now use the notation of quantum channels and quantum states, because it includes the case of stochastic maps on classical probability distributions. The following reasoning is therefore valid for both the classical and the quantum Fisher information matrix. Recall that the information matrix arises at the second order approximation when we perturb the parameters of a quantum state slightly:
\begin{align}
    d(\rho(\ttheta), \rho(\ttheta + \ddelta)) = \frac{1}{2}\ddelta^{\transp} M[\rho(\ttheta)] \ddelta + O(\lVert \ddelta \rVert^3).
\end{align}
The condition of monotonicity implies that the distance measure decreases if we apply some sort of quantum channel $\Phi$, that can represent a variety of different operations:
\begin{align}
    d(\Phi[\rho(\ttheta)], \Phi[\rho(\ttheta + \ddelta)]) \leq
    d(\rho(\ttheta), \rho(\ttheta + \ddelta)).
\end{align}
But this also must hold in the limit $\lVert \ddelta \rVert \to 0$, where we can drop higher order terms and only use the second order approximation of the distance. This means that
\begin{align}
    \frac{1}{2}\ddelta^{\transp} M[\Phi[\rho(\ttheta)]] \ddelta \leq \frac{1}{2}\ddelta^{\transp} M[\rho(\ttheta)] \ddelta.
\end{align}
We assumed that $\ddelta$ was a very short vector to derive this inequality, but now we can also rescale it again to arbitrary length. This means that the above inequality holds for \emph{any} $\ddelta$ and that it implies the \emph{matrix inequality} 
\begin{align}
    M[\Phi[\rho(\ttheta)]] \leq M[\rho(\ttheta)].
\end{align}
\begin{figure}
    \centering
    \includegraphics[width=.7\columnwidth]{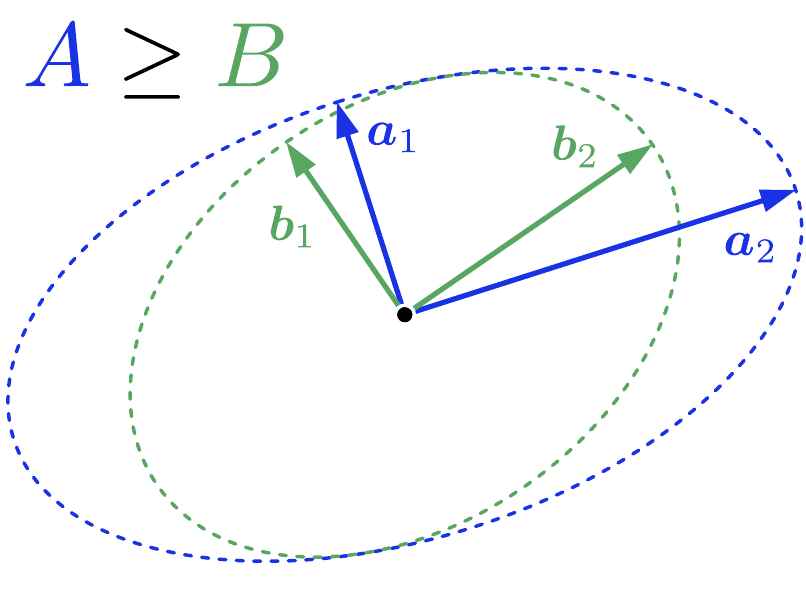}
    \caption{A visual explanation of the matrix inequality $A \geq B$ in the case of $2\times 2$ matrices. To every positive matrix, we can assign an ellipse by considering its action on the unit sphere -- in the 2D case the unit circle. The axes of the ellipse are then given by the eigenvectors scaled to the length of the associated eigenvector. The relation $A \geq B$ means that the ellipsis of $B$ lies inside the ellipsis of $A$. If the ellipses do not touch, we have the stronger relation $A > B$.}
    \label{fig:matrix_inequality}
\end{figure}
You might be a bit confused, as we are looking at matrices here and not numbers. The matrix inequality $A \geq B$ implies that the matrix $A - B$ has only non-negative eigenvalues, a statement that is equivalent to $\ddelta^{\transp} A \ddelta \geq \ddelta^{\transp} B \ddelta$ for any vector $\ddelta$. A more visual explanation of this is given in Fig.~\ref{fig:matrix_inequality}.

We have shown that the monotonicity of the distance measure means that the associated information matrix also has a monotonicity property as it can also only decrease under quantum operations. 
This fact helps us to shine light on the relation between the quantum Fisher information and the classical Fisher information. Remember that we already learned that we can also model \emph{measurements} as quantum channels. Together with the monotonicity of information matrices we just derived we know that applying a measurement $\calM$ need necessarily make the information matrix associated to any monotonic distance smaller: 
\begin{align}
    M[\calM[\rho(\ttheta)] \leq M[\rho(\ttheta)].
\end{align}
But the outcome of a measurement will always be a classical probability distribution over the measurement outcomes, ergo $\calM[\rho(\ttheta)]$ is a classical probability distribution. And due to the uniqueness of the classical Fisher information we therefore know that no matter what kind of distance we used to derive $M$, after the measurement it will be a constant multiple of the classical Fisher information matrix associated with the probability distribution after the measurement:
\begin{align}
    M[\calM[\rho(\ttheta)] = \alpha I[\calM[\rho(\ttheta)]].
\end{align}
As we are free to rescale the distance measure we use and therefore also the associated information matrix we can always make it consistent with the classical Fisher information when evaluated on classical probability distributions, which means that we can choose $\alpha = 1$, which we will also assume in the following. The relation is especially true for the quantum and the classical Fisher information, where we have that
\begin{align}
    I[\calM[\rho(\ttheta)]] \leq \calF[\rho(\ttheta)] \text{ for all }\calM.
\end{align}

Let us quickly take a step back and marvel at the feat we just accomplished: We have used the simple requirement of monotonicity of the distance measure to show that any quantum information matrix is an upper bound to the classical Fisher information matrix associated with \emph{any} probability distribution that results of a measurement of the state. We have initially motivated the derivation of this quantity via a geometric intuition, and it also pops up here: Indeed, a quantum information measure like the quantum Fisher information will measure how much the underlying quantum state $\rho(\ttheta)$ will change if we change the parameters of the state slightly. The relation we just derived tells us now that the change of the underlying quantum state directly gives us an upper bound on how much of this change we can make visible when performing a measurement.

This also gives us as a hint on what kinds of applications these two quantities enable. The quantum Fisher information really captures information about the underlying quantum state and therefore also phenomena that are quantum mechanical in nature. But the importance of the classical Fisher information should not be discounted, as we are always forced to perform measurements to extract information about the quantum system, which means that at the end of the day the classical Fisher information will always be the measure that quantifies the objects we can actually observe. This means that both quantities are very important for near-term applications, as the classical Fisher information captures the outputs of our experiments, but the quantum Fisher information can inform us about the quantum phenomena happening and also quantifies the ultimate limits of our approaches.

One might wonder if there always exists a measurement that achieves equality of the classical and the quantum Fisher information matrix. In the single-parameter case this is actually always possible~\cite{braunstein1994statistical}, and consequently we can always find a measurement that achieves the quantum Fisher information for every \emph{individual} parameter $\theta_i$. The optimal measurement can be found by computing the SLD operator $L_i$ that realizes the derivative of the underlying quantum state with respect to the parameter $\theta_i$ as in Eq.~\eqref{eqn:sld_operator_implicit_def}~\cite{liu2020quantum} and can depend on the actual value of $\ttheta$~\cite{barndorff-nielsen2000fisher}. 
But for multiple parameters we cannot necessarily find a measurement that achieves equality of classical and quantum Fisher information matrix, as the optimal measurements for the individual parameters need not be compatible with each other. A detailed discussion of the question of optimal measurements is found in Refs.~\cite{pezze2017optimal,yang2019optimal}. More detailed studies quantifying the (in)compatibility of different measurements can be found in Refs.~\cite{ragy2016compatibility,lu2021incorporating,belliardo2021incompatibility}.

\section{The Role of Noise}\label{sec:noise}
As is already evident from the name \emph{noisy} intermediate-scale quantum devices, noise is one of the principal impediments and a defining factor for NISQ devices. It is therefore imperative for us to understand how noise influences the classical and quantum Fisher information. We already got to know the concept of a quantum channel and learned that it can also be used to model any noisy quantum evolution. 

But up to now, we have only treated the quantum Fisher information for pure quantum states, resting on the particularly simple formula for the fidelity for pure states. If we wish to extend it to mixed quantum states, we have to find a quantity that reproduces the known fidelity formula for pure states but also stays valid for mixed quantum states. This generalization is given by the \emph{Bures fidelity}, also known as \emph{Uhlmann's fidelity}\footnote{Again, there exist different conventions if the square should be included or not.}
\begin{align}\label{eqn:def_bures_fidelity}
f_B(\rho, \sigma) = \Tr \{ (\rho^{1/2} \sigma \rho^{1/2})^{1/2} \}^2.
\end{align}
The notation $\rho^{1/2}$ denotes the unique positive \emph{matrix square root} of $\rho$, which is the only positive semidefinite matrix that fulfills $(\rho^{1/2})^2 = \rho$. If we look at the eigendecomposition of $\rho$, 
\begin{align}
    \rho = \sum_i \lambda_i | \lambda_i \rangle \! \langle \lambda_i |,
\end{align}
we can easily see that it is given by
\begin{align}
    \rho^{1/2} = \sum_i \sqrt{\lambda_i} | \lambda_i \rangle \! \langle \lambda_i |.
\end{align}
Constructions of the form $\rho^{1/2}\sigma \rho^{1/2}$ often appear in the context of quantum information theory because this is a way to multiply $\rho$ and $\sigma$ that -- contrary to just using $\rho \sigma$ -- yields a Hermitian matrix. 

As in the pure case, we need to transform the fidelity to get the proper distance measure associated to it, namely the \emph{Bures distance}
\begin{align}
    d_B(\rho, \sigma) = 2 - 2 f_B(\rho, \sigma).
\end{align}
The prefactor 2 ensures that the the associated information matrix is consistent with the classical Fisher information matrix, as in Sec.~\ref{sec:qfim}. 

We won't reproduce the whole derivation of the quantum Fisher information matrix for the Bures distance, as it is quite laborious. The interested reader can refer to Ref.~\cite{liu2014fidelity} for a detailed derivation. The resulting formula is:
\begin{align}\label{eqn:formula_qfim}
    \calF_{ij} &= \sum_{\substack{kl\\ \lambda_k + \lambda_l \neq 0}} \frac{2 \operatorname{Re}(\langle \lambda_k | \partial_i \rho | \lambda_l \rangle \! \langle \lambda_l | \partial_j \rho | \lambda_k \rangle)}{\lambda_k + \lambda_l} \\
    &= \sum_{\substack{k \\ \lambda_k \neq 0}} \frac{(\partial_i \lambda_k )(\partial_j \lambda_k)}{\lambda_k} + 4 \lambda_k \operatorname{Re}(\langle \partial_i \lambda_k | \partial_j \lambda_k \rangle) \nonumber\\
    &\hphantom{=}-\sum_{\substack{kl \\ \lambda_k, \lambda_l \neq 0}} \frac{8 \lambda_k \lambda_l}{\lambda_k + \lambda_l} \operatorname{Re}(\langle \partial_i \lambda_k | \lambda_l \rangle \! \langle \lambda_l | \partial_j \lambda_k \rangle).
\end{align}
Let us analyze this behemoth. First, we see that the equations contain a division by $\lambda_k + \lambda_l$. As $\rho$ is a positive semidefinite matrix, the eigenvalues $\lambda_k$ cannot be negative, which means that the case excluded in the sum can only occur if both $\lambda_k$ and $\lambda_l$ are zero. It is an advantage of the second equation that the sums now only run over $\lambda_k$ and $\lambda_l$ that are non-zero.

Now let's go over the two parts of the second equation. The first term,
\begin{align}
    \sum_{\substack{k \\ \lambda_k \neq 0}} \frac{(\partial_i \lambda_k )(\partial_j \lambda_k)}{\lambda_k},
\end{align}
looks familiar. Recall that in the eigenbasis, $\rho$ is diagonal and therefore represents a \emph{classical} probability distribution over the basis states with probabilities $\lambda_k$ -- this means that the term above is the \enquote{classical} part of the quantum Fisher information and quantifies how the eigenvalues themselves change. Note that here we only sum over $\lambda_k \neq 0$ -- which means that we effectively exclude the possibility of an $\lambda_k$ being zero but $\partial_i \lambda_k$ being nonzero. In such a case, the rank of the density matrix would change which causes the quantum Fisher information to be undefined~\cite{seveso2020discontinuity,safranek2017discontinuities}. This, however, is more of a technicality as there always exists a full-rank state arbitrarily close to any state we could be interested.

The next term now captures how much the \emph{eigenstates} themselves change under the parameters $\theta_i$ and $\theta_j$:
\begin{equation}
\begin{split}
    &\sum_{\substack{k \\ \lambda_k \neq 0}} 4 \lambda_k \operatorname{Re}(\langle \partial_i \lambda_k | \partial_j \lambda_k \rangle)\\
    &-\sum_{\substack{kl \\ \lambda_k, \lambda_l \neq 0}} \frac{8 \lambda_k \lambda_l}{\lambda_k + \lambda_l} \operatorname{Re}(\langle \partial_i \lambda_k | \lambda_l \rangle \! \langle \lambda_l | \partial_j \lambda_k \rangle).
\end{split}
\end{equation}
This constitutes the quantum part of the quantum Fisher information. The changing of the eigenvectors is a non-classical phenomenon as they are always fixed and identified with the different measurement outcomes for classical states. 

\paragraph{Calculation.}
We have seen that the noisy quantum Fisher information is much more complicated than the quantum Fisher information for pure states. To evaluate it exactly one usually has to perform a full tomography of the underlying state, an operation that is too costly for near-term applications because the number of samples is exponential in the number of qubits~\cite{baumgratz2013scalable}. For quantum states that are nearly pure approximations can be used, but they also break down at certain noise levels~\cite{toth2018lower,koczor2020quantum}. 

Recently, variational approaches for the computation of the Bures fidelity $f_B$ have been suggested. They can be used to compute the quantum Fisher information for noisy states in conjunction with the perturbation techniques described in Sec.~\ref{sec:qfim}. 
Ref.~\cite{du2021exploring} proposes to alleviate the resource requirements for tomography through the use of a variational quantum autoencoder. A variational autoencoder compresses an $N$-qubit input state into $K$ qubits~\cite{romero2017quantum}. The authors of Ref.~\cite{du2021exploring} show that the compressed state of a perfectly trained autoencoder has the same spectrum as the original state. They propose to estimate the Bures fidelity between two states $\rho$ and $\sigma$ by first performing tomography of the compressed state of $\rho$ and then running separate SWAP tests that involve the eigenstates of $\rho$ and the state $\sigma$. The fidelity is then computed in classical post-processing and guarantees on the estimation precision are given based on how well the autoencoder was trained. The authors provide numerical evidence that the proposed strategy works well for low-rank states if the variational circuit for the autoencoder is suitably chosen.

Ref.~\cite{chen2020variational}, on the other hand, proposed to exploit Uhlmann's theorem that states that the Bures fidelity of two states $\rho$ and $\sigma$ is equal to the maximum fidelity over all possible purifications $\ket{\psi_{\rho}}$ and $\ket{\psi_{\sigma}}$:
\begin{align}\label{eqn:uhlmanns_theorem}
    f_B(\rho, \sigma) = \max_{\ket{\psi_{\rho}},\ket{\psi_{\sigma}}} f(\ket{\psi_{\rho}}, \ket{\psi_{\sigma}}).
\end{align}
The authors suggest to use a variational circuits to learn purifications of the input states $\rho$ and $\sigma$ and then perform another optimization to extract the maximum in Eq.~\eqref{eqn:uhlmanns_theorem}.
Another approach for the calculation of the Bures fidelity was put forward in Ref.~\cite{tan2021variational}. It is based on a different subroutine, \emph{purity minimization}, that can be used to calculate the matrix square roots in Eq.~\eqref{eqn:def_bures_fidelity}. The two latter approaches require multiple copies of the input states and therefore incur an overhead that is large for NISQ applications. They furthermore rely on the success of variational subroutines which is closely tied to the successful selection of a circuit ansatz. These approaches are therefore only suitable to calculate the quantum Fisher information for small systems.

Ref.~\cite{sone2021generalized} introduced the \emph{truncated quantum Fisher information (TQFI)} as a way to approximate the quantum Fisher information by only including the quantum state's $m$ eigenvectors with the largest eigenvalues in the computation of the quantum Fisher information. For increasing $m$, the approximations obtained via the truncated approach get closer to the true quantum Fisher information. Recent efforts showed that the TQFI and other upper and lower bounds on the quantum Fisher information can be effectively evaluated on NISQ devices via variational procedures~\cite{beckey2020variational}.

The authors of Ref.~\cite{rath2021quantum} suggest another strategy in the same vein. They propose a hierarchy of lower bounds $\calF_n$ based on a series expansion of the quantum Fisher information that can be computed using randomized measurements~\cite{huang2020predicting}. The first level, $\calF_0$, already appeared in Refs.~\cite{zhang2017detecting,girolami2017witnessing} and represents a tighter bound than the sub-quantum Fisher information of Ref.~\cite{beckey2020variational}. With increasing $n$, the complexity of computing the lower bound increases, but the distance of the bound to the true quantum Fisher information decreases exponentially in $n$.

\section{NISQ Applications}\label{sec:nisq_applications}
We will now survey the variety of different contexts in which Fisher information popped up related to NISQ devices. The fact that these application are widely different shows the value of these techniques.

\subsection{Quantum Sensing}
Quantum sensing, also known as quantum metrology or quantum parameter estimation, is one of the main pillars of near-term quantum technologies. It is the area of application in which the classical and quantum Fisher information have been explored the most. We will explore quantum sensing to the degree necessary to understand the role of Fisher information, but, as the focus of this work is on near-term applications \emph{beyond} quantum sensing, we will not delve deeper into applications. But before we can do that, we first have to set the stage.

\paragraph{Introduction.}
The object of quantum sensing is to measure some physical quantity, say a magnetic field, pressure or temperature, which we will denote as $\pphi$. We consider a vector-valued quantity because we often want to measure multiple things, \emph{e.g.}\ all components of the magnetic field. The measurement proceeds by preparing a physical system that interacts with its environment so that the physical parameter is imprinted on it. Consider the example of a magnetic field: to find out what the components are we can use one or multiple spins that undergo precession depending on the strength of the magnetic field and the alignment of the spin relative to the magnetic field. 

In the end, we can model this whole process by taking a quantum system in an initial state $\rho_0$ which then undergoes a quantum channel $\calE(\pphi)$ that depends on the physical parameters. The result will again be a parametrized state:
\begin{align}
    \rho(\pphi) = \calE(\pphi)[\rho_0].
\end{align}
As the state $\rho_0$ \enquote{probes} the environment, it is called the \emph{probe state}. Note also that we are actually back in the framework of parametrized quantum states with which we started our discussion in Sec.~\ref{sec:preliminaries}, only that the parameters this time are the physical parameters we want to measure instead of some tunable parameters of the state preparation. 

The story is of course not over here. We want to use the obtained quantum state $\rho(\pphi)$ to learn as much as we can about the underlying parameters $\pphi$. To this end, we have to perform some measurement $\calM$. After the measurement, we are left with a probability distribution over measurement outcomes that depends on the physical parameters and the chosen measurement, $p_{\calM}(\pphi)$.
\begin{figure}
    \centering
    \includegraphics{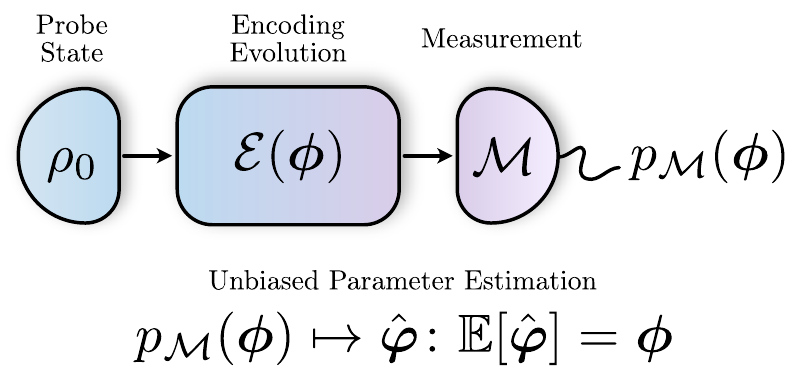}
    \caption{Mathematical representation of a quantum sensing experiment. A probe state $\rho_0$ undergoes an interaction with the environment that imprints the physical parameters $\pphi$ we want to sense onto the state. To extract information about the parameters we perform a measurement $\calM$ of the final state, yielding a probability distribution $p_{\calM}(\pphi)$ that depends on both the chosen measurement and the physical parameters. From this probability distribution, an estimator of the physical parameters, $\hat{\vvarphi}$ is constructed.}
    \label{fig:metrology_pipeline}
\end{figure}
If we make an estimate of the underlying parameters from the observed measurement statistics we formally construct an \emph{estimator}, denoted as $\hat{\vvarphi}$. As the measurement results are random, our estimate will necessarily also be and the estimator $\hat{\vvarphi}$ is therefore a \emph{random vector}. A property of an estimator that we can aim for is that it is \emph{unbiased}, \emph{i.e.}\ that the predictions are correct in expectation: $\bbE[\hat{\vvarphi}] = \pphi$. There are also other notions of \enquote{unbiasedness} that are less restrictive: An estimator can be \emph{locally unbiased}, which means it is only unbiased in the neighborhood of a certain parameter value $\pphi_0$. This property is of interest if one already has prior knowledge about the underlying parameter, \emph{e.g.}\ from previous measurements. Furthermore, an estimator can be \emph{asymptotically unbiased}, \emph{i.e.}\ it is unbiased in the limit of infinitely many samples.
The whole formalization of a sensing experiment is depicted in Fig.~\ref{fig:metrology_pipeline}.

\paragraph{Quantifying Sensing Performance.}
The classical Fisher information matrix comes into play when we want to quantify how good an estimator can be if we perform our experiment $n$ times. This manifests itself in the \emph{Cramér-Rao bound}, which is central to the field of quantum sensing and quantifies the best attainable performance of an unbiased estimator:
\begin{align}\label{eqn:cramer_rao_bound}
    \operatorname{Cov}[\hat{\vvarphi}] \geq \frac{1}{n} I_{\calM}(\pphi)^{-1}.
\end{align}
This is again a matrix inequality, as explained in Fig.~\ref{fig:matrix_inequality}. Let us decode this inequality. On the left hand side there is the \emph{covariance matrix} of our estimator, whose entries are the covariances of the separate components
\begin{align}
    \operatorname{Cov}[\hat{\vvarphi}]_{ij} = \bbE[\hat{\varphi}_i \hat{\varphi}_j] - \bbE[\hat{\varphi}_i] \bbE[\hat{\varphi}_j].
\end{align}
As we want to be sure of our estimates, we want our estimator to vary as little as possible, which means that the covariances should be as small as possible. We can also connect the covariance matrix to the \emph{mean-squared error} of our estimate
\begin{align}\label{eqn:mse_cov_matrix}
    \operatorname{MSE}[\hat{\vvarphi}]
    = \bbE [ \lVert \hat{\vvarphi} - \pphi \rVert^2 ]
    = \Tr\{\operatorname{Cov}[\hat{\vvarphi}]\},
\end{align}
which can be more easily interpreted as a performance measure. Note that this identity only holds for unbiased estimators. 

The right hand side of the Cramér-Rao bound Eq.~\eqref{eqn:cramer_rao_bound} imposes a fundamental limit on how small the covariances can get. And here we find the inverse of the classical Fisher information matrix $I_{\calM}(\pphi)$. Note that we denoted the chosen measurement explicitly, because the classical Fisher information matrices for different measurements usually do not coincide. 

The appearance of the Fisher information has an intuitive explanation: Remember that large entries of the Fisher information indicate that a change of the associated parameters results in a large change of the underlying probability distribution. This would be desirable for the purpose of estimation because this means that a probability distribution we observe can be associated with a certain parameter with higher confidence -- if the parameter was different the probability distribution would also be notably different. As the Cramér-Rao bound is concerned with the covariance, a quantity that we want to be as small as possible, we need the \emph{inverse} of the Fisher information matrix to account for that.

A property that makes the scalar Cramér-Rao bound very useful is that it can \emph{always} be saturated in the limit of infinitely many samples. In this limit, a process called maximum likelihood estimation is guaranteed to give an optimal estimate. This is appealing, because it means that we can stop worrying about how to actually construct a good estimator because we know there must be one that achieves the Cramér-Rao bound.

\paragraph{Optimal Sensing.}
We see that a sensing procedure has two levers we can pull to optimize it: we want to choose a probe state that is \enquote{maximally susceptible} to the evolution $\calE(\pphi)$ and we want to find a measurement $\calM$ that extracts as much of this information from the quantum state. 

We learned in Sec.~\ref{sec:relation_cfim_qfim} that the quantum Fisher information matrix gives an upper bound to the classical Fisher information matrices that can be obtained by measuring the underlying quantum system. This upper bound on the classical Fisher information matrix directly implies a more fundamental lower bound on the attainable precision, the so-called \emph{quantum Cramér-Rao bound}~\cite{helstrom1969quantum,holevo2011probabilistic}
\begin{align}\label{eqn:quantum_cramer_rao_bound}
    \operatorname{Cov}[\hat{\vvarphi}] \geq \frac{1}{n} I_{\calM}(\pphi)^{-1}\geq \frac{1}{n} \calF(\pphi)^{-1}.
\end{align}
The quantum Cramér-Rao bound has the advantage that it is independent of the chosen measurement and only depends on the probe state. We can therefore formulate an intuitive recipe to find an optimal quantum sensing scheme: first try to find the probe state with the largest quantum Fisher information matrix and then optimize the POVM to bring the classical Fisher information matrix as close to the quantum one as possible.

As pointed out in Sec.~\ref{sec:relation_cfim_qfim}, we are often unable to construct a measurement for which $I_{\calM} = \calF$ due to the possible incompatibility of the parameters $\phi_i$. But how can we then decide which possible measurement is \enquote{best}?
We have just learned that we can construct an estimator that achieves the classical Cramér-Rao bound for any measurement. But we can usually not use the covariance matrix to quantitatively compare the performance of two estimators $\hat{\vvarphi}_1$ and $\hat{\vvarphi}_2$ corresponding to different measurements. This is because it can be that neither $\operatorname{Cov}[\hat{\vvarphi}_1] \geq \operatorname{Cov}[\hat{\vvarphi}_2]$ nor $\operatorname{Cov}[\hat{\vvarphi}_2] \geq \operatorname{Cov}[\hat{\vvarphi}_1]$. This can be intuitively understood by looking at Fig.~\ref{fig:matrix_inequality}: if neither ellipse corresponding to the two covariance matrices is contained in the other, we cannot make a statement of one being \enquote{larger} than the other.  

To still be able to make a comparison we need a scalar quantity. To get one, we can generalize the idea of Eq.~\eqref{eqn:mse_cov_matrix} and perform a trace of the covariance matrix with the addition of a positive semidefinite \emph{weight matrix} $W$. As the name suggests, we can use it to put additional emphasis on certain parameters or just use the identity matrix to perform an equal weighting. In this case, we get a scalar quantity that quantifies the performance of an estimator and that fulfills a scalar (quantum) Cramér-Rao bound: 
\begin{align}\label{eqn:weighted_cramer_rao_bound}
\begin{split}
    \Tr\{ W \operatorname{Cov}[\hat{\vvarphi}] \} &\geq \frac{1}{n} \Tr\{W I_{\calM}(\pphi)^{-1}\} \\ &\geq \frac{1}{n} \Tr\{W \calF(\pphi)^{-1}\}.
\end{split}
\end{align}

We should note that the quantum Cramér-Rao bound is not the end of the story. If we consider a weighting of the estimator's covariance matrix as in Eq.~\eqref{eqn:weighted_cramer_rao_bound}, the \emph{Holevo Cramér-Rao bound (HCRB)} gives the ultimate limit to estimation precision and is attainable in the limit of infinitely many samples~\cite{demkowicz-dobrzanski2020multi-parameter}. It is computed by solving an optimization problem and does not involve the quantum Fisher information matrix. It can, however, not exceed the quantum Cramér-Rao by more than a factor of 2~\cite{tsang2020quantum} and is equal to it when the underlying parameters can be estimated simultaneously. 

For a review on multi-parameter quantum sensing, have a look at Refs.~\cite{albarelli2020perspective,sidhu2020geometric,liu2020quantum}. Ref.~\cite{sidhu2020geometric} that provides an exhaustive review with a particular focus on the underlying geometry.

\paragraph{Exploiting Quantum Effects.}
The (quantum) Cramér-Rao bound contains a factor $1/n$ which accounts for the fact that we can simply decrease the variance of our estimates by averaging over $n$ independent repetitions of the same experiment. The scaling
\begin{align}
    \operatorname{Cov}[\hat{\vvarphi}] \propto \frac{1}{n}
\end{align}
is called the \emph{standard quantum limit (SQL)} or \emph{shot-noise limit}.

But the approach of just performing independent repetitions of the same experiment does not make use of one of the most crucial properties of quantum mechanics, namely \emph{entanglement}. To elucidate how we can make use of it, we have a look at a simple sensing task where we want to measure the rate $\Delta$ at which a qubit acquires a phase. In our experiment every single-qubit probe acquires a phase $\phi = \Delta t$ between the $\ket{0}$ and the $\ket{1}$ state. If we perform $n$ repetitions separately, we obtain the scaling of the standard quantum limit. But we can also entangle the $n$ probes into a generalized GHZ state: 
\begin{align}
\ket{\text{GHZ}_n} &= \frac{1}{\sqrt{2}} \ket{\underbrace{00\dots 0}_{n\text{ times}}} +  \frac{1}{\sqrt{2}} \ket{\underbrace{11\dots 1}_{n\text{ times}}} \\
&= \frac{1}{\sqrt{2}}\ket{0_n} + \frac{1}{\sqrt{2}}\ket{1_n}.
\end{align}
Under the sensing interaction, the state evolves to:
\begin{align}
\frac{1}{\sqrt{2}} \ket{0_n} + e^{-i n \Delta t} \frac{1}{\sqrt{2}} \ket{1_n}.
\end{align}
Effectively, we have enhanced the signal by a factor of $n$ compared 
to a single probe.

This factor of $n$ enhancement lies at the heart of quantum advantage in sensing. You might rightfully ask why this is the case, because we also got a factor $n$ enhancement from just performing $n$ separate repetitions. But it turns out that the fact that we enhanced the \emph{signal} by a factor of $n$ actually gives us a factor $n^2$ improvement in the Fisher information! Recall that the classical Fisher information with respect to the parameter $\Delta$ is given by
\begin{align}
    I(\Delta) = \sum_l \frac{(\partial_{\Delta} p_l)^2}{p_l}.
\end{align}
By enhancing the signal by a factor of $n$ we actually changed the rate with which it changes -- the derivative with respect to $\Delta$ -- by a factor of $n$ compared to a single repetition. Because the Fisher information contains the \emph{square} of the derivative, this effectively gives us the enhancement by a factor of $n^2$.

The scaling of this entangled approach,
\begin{align}
    \operatorname{Cov}[\hat{\vvarphi}] \propto \frac{1}{n^2},
\end{align}
is actually the most fundamental limit attainable when exploiting quantum mechanical effects and is called the \emph{Heisenberg limit}~\cite{bollinger1996optimal}. 

But there is a catch. If we consider realistic sensing problems with noise we see that the generalized GHZ state we just used to achieve Heisenberg scaling is actually no better than the simple single-qubit strategy~\cite{huelga1997improvement}, even if there is only an infinitesimal amount of noise. This is due to the fact that not only the signal but also the noise itself gets \enquote{amplified}, therefore canceling out the advantage from quantum entanglement. But not all hope is lost as Ref.~\cite{huelga1997improvement} also showed that other, less entangled states, can still give an advantage in the noisy regime. More recently, it was shown in Ref.~\cite{demkowicz-dobrzanski2012elusive} that one can only hope to get a constant factor improvement to the standard quantum limit for many noise models. To regain the quantum advantage, one needs to combine quantum error correction with quantum sensing. Indeed, if the noise acts sufficiently different than the signal, we can retain the Heisenberg scaling by using a \emph{metrological code}~\cite{zhou2018achieving,gorecki2020optimal}. For certain noise models, scalings between standard and Heisenberg scaling can be achieved~\cite{haase2016precision}.

\paragraph{Estimation of Expectation Values.}
Estimating expectation values is a key subroutine of variational quantum algorithms. Recently, the authors of Refs.~\cite{wang2021minimizing,koh2020framework} used reasoning from quantum sensing to find a way to reduce the number of evaluations of a NISQ experiment necessary to faithfully evaluate an expectation values at the cost of increased circuit depth.

The approach is based on the observation that the estimation quality of an expectation value of a Pauli observable $\langle P \rangle$ is also governed by the Cramér-Rao bound. We can use the same constructions as we have already outlined but identify the unknown parameter with the expectation value: $\phi = \langle P \rangle$. If we want to achieve a fixed precision we can reduce the number of experimental repetitions necessary to achieve it by increasing the classical Fisher information with respect to the parameter $\langle P \rangle$. The construction of Refs.~\cite{wang2021minimizing,koh2020framework} is enabled by the use of a Bayesian approach, where an initial estimate about the distribution of $\langle P \rangle$ is updated with new information from experiments. The classical Fisher information is increased by a subroutine inspired from Grover's famous algorithm. The authors call this approach \emph{Engineered Likelihood Functions}, as the subroutines inspired by Grover's algorithm have parameters that are adjusted to yield the highest possible Fisher information for the specific expectation value that shall be estimated.

\paragraph{Variational Algorithms for Sensing.}
As metrology is such a fundamental application, it is important to construct optimal protocols. This is rather difficult because noise and device limitations have to be taken into account. But recently the idea of using NISQ devices themselves for the optimization of quantum sensing schemes has gained some attention and a number of variational algorithms for that purpose have been put forward.

In the above exposition of quantum sensing, we looked only at the physical parameters $\pphi$ as parameters of the state. But nothing holds us back from parametrizing the probe state as well, maybe in a simulation on a NISQ computer or by providing a tunable state preparation in an experiment. The same also holds for the measurement -- we can fix a certain detection scheme but precede it by a unitary, again with parametrized gates. If we denote the parameters of the probe state preparation with $\ttheta$ and the parameters of the measurement with $\mmu$, we work with a parametrized state $\rho(\ttheta, \pphi, \mmu)$. 

To perform a variational algorithm we also need a \emph{cost function} that measures how well we do with our estimation. The first works in this direction considered the case of estimating only a single parameter. This line of work was started in Ref.~\cite{kaubruegger2019variational}, which considered the probe state's \emph{spin squeezing} as a cost function, as it acts as surrogate for the sensing precision. Subsequently, the authors of Ref.~\cite{koczor2020variational-state} used the change of the fidelity under small perturbation of the sensing parameter to estimate and optimize the quantum Fisher information, a method outlined in detail in Sec.~\ref{sec:qfim}. Note that looking at the quantum Fisher information is interesting because it captures the best achievable performance of the sensing scheme, but it also removes the performed measurement from the equation, as the quantum Fisher information is independent of $\mmu$. The authors of Ref.~\cite{meyer2020variational} argue that this will not be desirable in realistic applications where measurement capabilities are limited and the optimal measurement cannot be realized. Instead, they propose to use a cost function based on the classical Fisher information and extend the proposal to multi-parameter estimation. This technique was further improved in Ref.~\cite{ma2020adaptive} by combining it with an additional optimization over the parametrizations of the probe state and the measurement.

Another strategy to tackle this problem was put forward in Ref.~\cite{beckey2020variational}, where efficient schemes to evaluate different bounds on the quantum Fisher information were proposed and applied to the optimization of quantum sensing protocols. In a follow-up work, it was proven that the global optima of a specific bound on the quantum Fisher, the \emph{sub-quantum Fisher information}, first derived in Ref.~\cite{garttner2018relating}, coincide with that of the quantum Fisher information itself~\cite{cerezo2021sub-quantum}, underscoring its quality as a surrogate for the quantum Fisher information.

\subsection{Quantum Natural Gradient Descent}
Another intriguing application of Fisher information can be found in the field of variational quantum algorithms. In these applications, we desire to minimize a cost function $C$ that depends on some expectation values evaluated on a parametrized quantum state $\ket{\psi(\ttheta)}$, rendering the cost function a function of the circuit parameters, $C = C(\ttheta)$.

A popular way to minimize the cost function is by using \emph{gradient descent}. In this scheme, we start from an initial guess of parameters $\ttheta^{(0)}$ and perform multiple optimization steps where we update $\ttheta$ according to the rule
\begin{align}\label{eqn:qng_step}
    \ttheta^{(t+1)} = \ttheta^{(t)} - \eta \nabla C(\ttheta^{(t)}).
\end{align}
The gradient of $C(\ttheta)$ always points in the direction of the steepest increase of the cost function. Subtracting it therefore ensures we go into a direction where the cost function decreases. The parameter $\eta$ is the \emph{step size} that controls how far we step. 

The authors of Ref.~\cite{stokes2020quantum} proposed a way to make use of the knowledge about the underlying quantum states encoded in the quantum Fisher information matrix $\calF$ in the context of optimization, namely \emph{quantum natural gradient descent}. In this approach, the gradient descent update rule is modified by the use of the inverse of the quantum Fisher information matrix:
\begin{align}
    \ttheta^{(t+1)} = \ttheta^{(t)} - \eta  \calF(\ttheta)^{-1} \nabla C(\ttheta^{(t)}).
\end{align}
Let us gather some intuitive understanding of this approach. As we have seen in Eq.~\eqref{eqn:def_pullback_metric}, the fidelity distance is given by
\begin{align}
    d_f(\ttheta, \ttheta + \ddelta) = \frac{1}{4}\ddelta^{\transp} \calF \ddelta 
\end{align}
in leading order, see also Sec.~\ref{sec:qfim}. This means that large entries of the quantum Fisher information matrix correspond to large changes in the distance and small entries to small changes. By taking the inverse of the quantum Fisher information matrix we thus \emph{normalize} the gradient step we take: directions that correspond to large changes of the underlying quantum state are scaled down while directions that correspond to small changes are scaled up.

But we can also give a beautiful mathematical justification of this approach. This can be seen by analyzing the origin of the original gradient descent rule. We can recast the next set of parameters $\ttheta^{(t+1)}$ as the solution to the following optimization problem:
\begin{align}
    \operatornamewithlimits{argmin}_{\ttheta} \left\{ \langle \ttheta - \ttheta^{(t)}, \nabla  C(\ttheta^{(t)}) \rangle 
     + \frac{1}{2\eta}\lVert \ttheta - \ttheta^{(t)} \rVert_2^2  \right\}.
\end{align}
This includes a term that ensures we go in the opposite direction of the gradient and a so-called \emph{regularization term} that ensures that we do not step too far, thereby controlling the step size. 

But we have already learned that we can use the quantum Fisher information matrix to measure lengths of vectors, thereby including our knowledge about the underlying quantum state. We thus replace the 2-norm in the regularization term with the norm induced by $\calF$, $\lVert \ttheta \rVert_{\calF}^2 = \ttheta^{\transp} \calF \ttheta$ -- which is for small step sizes approximately equal to the expected fidelity distance $d_f$ up to a constant factor. This means that we do no longer penalize large distances in parameter space, but instead in the space of quantum states! The solution to the new optimization problem
\begin{align}
    \operatornamewithlimits{argmin}_{\ttheta} \left\{ \langle \ttheta - \ttheta^{(t)}, \nabla  C(\ttheta^{(t)}) \rangle + \frac{1}{2\eta}\lVert \ttheta - \ttheta^{(t)} \rVert_{\calF}^2  \right\}
\end{align}
is nothing but the update step of quantum natural gradient descent from Eq.~\eqref{eqn:qng_step}.

It was shown in Ref.~\cite{wierichs2020avoiding} that this method actually provides an advantage in optimizing realistic quantum systems when large systems are concerned due to it taking different optimization paths than optimization strategies that do not relate to the underlying quantum states. 

A weak point of quantum natural gradient descent is that the quantum Fisher information matrix has to be calculated at each step, which can be a very costly endeavor, as also outlined in Sec.~\ref{sec:qfim}. The authors of Ref.~\cite{van_straaten2020measurement}, however, argue that the additional cost of estimating the quantum Fisher information matrix is negligible in gradient-descent applications. This is because the number of shots necessary to faithfully evaluate the gradient of the cost function itself increases as one approaches a minimum whereas the cost of estimating the quantum Fisher information does not.

To reduce the complexity one can also resort to stochastic approximations of the quantum Fisher information matrix as proposed in Ref.~\cite{gacon2021simultaneous}. The authors use a variation of SPSA to approximate the quantum Fisher information matrix, as was already introduced in Sec.~\ref{sec:qfim}. The authors also use the assumption that the quantum Fisher information matrix only changes slowly between iterations to perform a smooth approximation of the quantum Fisher information matrix. In this approach, the approximation of the matrix is held in memory and the new stochastic approximations obtained at the current optimization step are used to update this approximation. This approach has the downside that it is not guaranteed that this approximation of the quantum Fisher information matrix has only positive eigenvalues, but the authors introduce measures to mitigate this issue.

Another issue with quantum natural gradient descent arises when we consider noisy quantum devices. As we already learned in Sec.~\ref{sec:noise}, the true quantum Fisher information is very hard to calculate for a noisy quantum state. The authors of Ref.~\cite{koczor2020quantum} argue that at least in the case of states that are not too noisy, approximations to the quantum Fisher information matrix are sufficient.

\subsection{Analyzing Quantum Learning Models}
\emph{Quantum Machine Learning (QML)}~\cite{schuld2015introduction,biamonte2017quantum} is a research field with increasingly growing traction. The field searches both for applications of quantum computers to improve machine learning techniques, as well as applications of machine learning to analyze quantum systems. A special focus lies on variational learning techniques useful in NISQ applications.

The classical Fisher information has already been successfully used in classical machine learning to analyze learning models like neural networks. A particular example is the \emph{Fisher-Rao norm}~\cite{liang2019fisher-rao}, which is nothing else than the norm induced by the classical Fisher information matrix that we already encountered:
\begin{align}
    \lVert \ddelta \rVert_{\mathrm{fr}}^2 = \ddelta^{\transp} I \ddelta.
\end{align}
The Fisher-Rao norm is treated as a measure that correlates with the \emph{capacity} of a learning model, which measures how complicated the relationships a learning model can express are. 

The authors of Ref.~\cite{abbas2020power} have looked at the classical Fisher information of a parametrized quantum circuit to quantify its capacity~\cite{abbas2020power}. They define a new capacity measure they call the \emph{effective dimension} which can be used to bound how well a variational quantum learning model can generalize on unseen data. The effective dimension also takes into account how many datapoints are available to train a learning model. In the limit of infinitely many datapoints, the effective dimension is equal to the number of non-zero eigenvalues -- the rank -- of the classical Fisher information matrix. A higher effective dimension is associated with a \enquote{flatter} eigenvalue spectrum of the classical Fisher information matrix. The authors also provide numerics that suggest that this \enquote{flatness} is a generic feature of certain quantum learning models.

Recently, the authors of Ref.~\cite{haug2021capacity} introduced the \emph{effective quantum dimension} as a measure of how much of the underlying quantum state space a quantum learning model can explore. The effective quantum dimension as proposed in this work is equal to the rank of the quantum Fisher information matrix. It does not take into account the number of available datapoints and should therefore not be confused with the quantum generalization of the effective dimension of Ref.~\cite{abbas2020power}. 
It is still an intuitive measure as the rank directly captures in how many directions a varying of the parameters will also result in a varying of the underlying quantum state. If the effective quantum dimension is lower than the number of parameters, then some of the parameters are redundant and the model is therefore overparametrized. The numerical investigations of the authors confirm that the specific choice of the circuit parametrization leads to different effective quantum dimensions of the resulting quantum learning models.

\section{Beyond NISQ}\label{sec:beyond_nisq}
The focus of this work is to showcase and explore the application of classical and quantum Fisher information in the NISQ context. But these tools have of course very interesting applications beyond that. We want to give two notable mentions that display how versatile these tools are:

In Ref.~\cite{yang2020optimal}, techniques from quantum sensing have been used to prove how large \enquote{quantum programs} need to be to allow a quantum computer to perform a unitary with given target precision. They construct an optimal quantum program interpreter that basically estimates the unitary that should be performed from the quantum program, therefore putting it in the realm where the tools from quantum metrology can be used to prove that the approach matches a lower bound.

Another very interesting application arose recently when other techniques from quantum metrology were used to to provide a new and simple proof for the approximate Eastin-Knill theorem in quantum error correction~\cite{kubica2020using}. The theorem shows that certain classes of error correcting codes with very favorable properties cannot exist. The new proof uses upper bounds on the quantum Fisher information to show that an error correcting code that violates the Eastin-Knill theorem would allow for too large Fisher information.

Recently, Ref.~\cite{tan2021fisher} showed that the classical and quantum Fisher information can be used in the framework of quantum resource theories~\cite{chitambar2019quantum}. The authors propose a general way to construct a parameter estimation task for which the presence of a resource gives an advantage. As this construction is generic, it also allows for the converse reasoning: every quantum resource is useful for metrology, because there exists a parameter estimation task for which it provides an advantage.

It is also worth noting that the review by Liu \emph{et al.}~\cite{liu2020quantum} contains a chapter of other applications of the quantum Fisher information in the contexts of quantum thermodynamics, quantum speed limits and the study of non-Markovianity. 

\section{Outlook}\label{sec:outlook}
Both the classical and the quantum Fisher information capture important information about the parametrized quantum systems that lie at the heart of many applications of near-term quantum devices. They can therefore further our understanding of the capabilities of these techniques and guide us in the development of new approaches. We hope that the present work will motivate some of its readers to incorporate the classical and quantum Fisher information into their practical and theoretical toolboxes and that it can inspire new uses of these tools in the context of NISQ devices.

As discussed in the main text, there have been many developments aiming to simplify the calculation of both the classical and quantum Fisher information on NISQ hardware. To further increase the applicability, it will, however, be necessary to further improve on these techniques. 

While the classical Fisher information is in principle easily accessible from the output probability distributions, there is still a need for rigorous analyses of different estimation strategies. A promising avenue, especially for variational applications, would be to consider Bayesian techniques, as the classical Fisher information is expected not to change too rapidly if circuit parameters are updated incrementally. Another important development would be rigorous bounds on the number of samples needed to faithfully estimate the classical Fisher information that could guide the computation in practical applications.

The quantum Fisher information on the other hand suffers from problems related to its higher complexity. An especially impactful development would be a technique that allows for the efficient calculation of the quantum Fisher information for noisy quantum states that are encountered in relevant settings. Machine learning approaches that have recently been developed to capture the properties of quantum states~\cite{torlai2020machine-learning} could possibly be useful in this regard. 

As classical and quantum Fisher information are so intimately tied to the structure of parametrized quantum states, we expect them to be useful in many more applications than discussed in this paper. 

An interesting direction is to further understand the properties of learning models that are constructed from parametrized quantum circuits through the lens of quantum Fisher information. Especially generalization bounds that are \enquote{truly quantum} would be of high interest, as they could capture the ultimate limits of quantum-enhanced machine learning models and lead to a better understanding of the influence of noise. It would be especially important in that regard to derive bounds that take into account the dichotomy between circuit parameters that represent data inputs and circuit parameters that are trainable.

It would furthermore be intriguing to see if more tools that were developed in the context of quantum sensing, \emph{e.g.}\ measures of parameter incompatibility, could be applied to the analysis of near-term applications. As variational quantum algorithms usually perform measurements in many bases to estimate expectation values of relevant observables, one possible directions would be to analyze the difference between classical and quantum Fisher information as this quantifies the information loss associated with specific measurements.

The fact that both classical and quantum Fisher information are susceptible to noise also suggests that they could be applicable to near-term quantum error correction and error mitigation, both in practical applications as well as theoretical tools to prove rigorous mathematical statements.

Another alluring avenue of research is to develop applications of monotone metrics beyond the quantum Fisher information, like the Wigner-Yanase information or the Kubo-Mori information. While these quantities do not provide tighter bounds in the quantum Cramér-Rao bound, they can give tighter bounds in other contexts~\cite{pires2016generalized}. It is natural to ask if generalizations of the approaches already developed with the quantum Fisher information in mind, \emph{e.g.}\ the quantum natural gradient descent, also perform well when it is replaced with a different monotone metric. An important prerequisite for such applications would be to find efficient ways to calculate those other metrics in practical applications.

\section*{Acknowledgments}
The author would like to thank Jacob Beckey, Francesco Albarelli, Daniel Stilck-França, Nathan Killoran and Josh Izaac for providing feedback that greatly improved the quality of the manuscript.

The author acknowledges funding from the German Federal Ministry for Economic Affairs and Energy under the PlanQK initiative.

\printbibliography

\clearpage
\appendix
\onecolumngrid

\section{Derivation of the Classical Fisher Information}\label{app:deriv_cfim}
We start out derivation of the classical Fisher information matrix from the Kullback-Leibler- or KL-divergence between a probability distribution and its perturbed counterpart
\begin{align}
    d_{\KL}(p_{\calM}(\ttheta), p_{\calM}(\ttheta + \ddelta)) = \sum_{l \in \calM} p_l( \ttheta) \log \frac{p_l(\ttheta)}{p_l(\ttheta + \ddelta)}.
\end{align}
We can exploit the fact that $\log(a/b) = \log a - \log b$ and rewrite the perturbed KL divergence as
\begin{align}\label{eqn:kl_div_rewrite}
    d_{\KL}(p_{\calM}(\ttheta), p_{\calM}(\ttheta + \ddelta)) = \sum_{l \in \calM} p_l( \ttheta) (\log p_l(\ttheta) - \log p_l(\ttheta + \ddelta)).
\end{align}
We will now compute the metric by performing a second order expansion around $\ddelta = 0$. To do so, we take the second derivatives with respect to the components of $\ddelta$, so it is immediately clear that only the second term of Eq.~\eqref{eqn:kl_div_rewrite} will contribute. We thus have
\begin{align}
    [M_{\KL}]_{ij} &= -\left.\frac{\partial^2}{\partial \delta_i \partial \delta_j}\sum_{l \in \calM} p_l(\ttheta)\log p_l(\ttheta + \ddelta)\right|_{\ddelta = 0} \\
    &= - \sum_{l \in \calM} p_l(\ttheta) \left.\frac{\partial^2 }{\partial \delta_i \partial \delta_j} \log p_l(\ttheta + \ddelta)\right|_{\ddelta = 0} \\
    &= - \bbE \left\{ \left.\frac{\partial^2 }{\partial \delta_i \partial \delta_j} \log p_l(\ttheta + \ddelta)\right|_{\ddelta = 0}\right\}.
\end{align}
We can now substitute $\xxi = \ttheta+ \ddelta$, in which case the derivatives transform as 
\begin{align}
    \frac{\partial}{\partial \delta_i} = \frac{\partial}{\partial \xi_i}\frac{\partial \xi_i}{\partial \delta_i} = \frac{\partial }{\partial \xi_i}
\end{align}
to obtain
\begin{align}
    [M_{\KL}]_{ij} &= \bbE \left\{\left. - \frac{\partial^2 }{\partial \xi_i \partial \xi_j}\log p_l(\xxi) \right|_{\xxi = \ttheta}\right\}.
\end{align}
Now we can rename $\xxi$ to $\ttheta$ again to get the more familiar looking
\begin{align}
    [M_{\KL}]_{ij} &= \bbE \left\{ - \frac{\partial^2 }{\partial \theta_i \partial \theta_j}\log p_l(\ttheta) \right\}.
\end{align}
We can also continue this derivation a bit more to get a second form of this expression. Consider that
\begin{align}
    -\frac{\partial^2 }{\partial \theta_i \partial \theta_j}\log p_l(\ttheta)  
    =  -\frac{\partial }{\partial \theta_j} \frac{1}{p_l(\ttheta)} \frac{\partial p_l(\ttheta)}{\partial \theta_i}
    =  \frac{1}{p_l^2(\ttheta)} \frac{\partial p_l(\ttheta)}{\partial \theta_j}\frac{\partial p_l(\ttheta)}{\partial \theta_i} - \frac{1}{p_l(\ttheta)} \frac{\partial^2 p_l(\ttheta)}{\partial \theta_i\partial \theta_j}.
\end{align}
To get to the expression for $[M_{\KL}]_{ij}$, we have to take the expectation value of this expression over the whole probability distribution. In the process of doing so, the second term will actually disappear:
\begin{align}
    \sum_{l \in \calM } p_l(\ttheta) \frac{1}{p_l(\ttheta)} \frac{\partial^2 p_l(\ttheta)}{\partial \theta_i\partial \theta_j}  &=
    \sum_{l \in \calM } \frac{\partial^2 p_l(\ttheta)}{\partial \theta_i\partial \theta_j} \\
    &=  \frac{\partial^2 }{\partial \theta_i\partial \theta_j} \sum_{l \in \calM} p_l(\ttheta) \\
    &=   \frac{\partial^2 }{\partial \theta_i\partial \theta_j} 1 \\
    &=   0.
\end{align}
To summarize, we arrive at the following equivalent formulas:
\begin{align}
    [M_{\KL}]_{ij} &=  \sum_{l \in \calM }p_l(\ttheta)   \frac{\partial^2 }{\partial \theta_i \partial \theta_j}\log p_l(\ttheta) \\
    &= \sum_{l \in \calM } \frac{1}{p_l(\ttheta)} \frac{\partial p_l(\ttheta)}{\partial \theta_j}\frac{\partial p_l(\ttheta)}{\partial \theta_i}.
\end{align}

\section{Properties of the Classical Fisher Information Matrix}\label{app:properties_cfim}
The classical Fisher Information Matrix $I$ of a probability distribution $p(\ttheta)$ with respect to a set of $d$ parameters $\ttheta \in \bbR^{d}$ is given by
\begin{align}\label{eqn:formula_cfim_appendix}
    I_{ij} &=  \sum_{l}p_l(\ttheta)   \frac{\partial^2 }{\partial \theta_i \partial \theta_j}\log p_l(\ttheta) = \sum_{l} \frac{1}{p_l(\ttheta)} \frac{\partial p_l(\ttheta)}{\partial \theta_i}\frac{\partial p_l(\ttheta)}{\partial \theta_j}
\end{align}
where the index $l$ iterates over all elements of $p(\ttheta)$. It has the following properties:
\begin{enumerate}[(i)]
    \item $I$ is a real symmetric $d\times d$ matrix.
    \begin{align}
        I \in \bbR^{d\times d} \qquad I_{ij} = I_{ji}
    \end{align}
    
    \item $I$ is positive semidefinite, \emph{i.e.}\ it only has non-negative eigenvalues. We write this as
    \begin{align}
        I \geq 0.
    \end{align}
    
    \item Convexity. $I$ is convex, which means that for any two probability distributions $p(\ttheta)$ and $q(\ttheta)$,
    \begin{align}
        I[\lambda p(\ttheta) + (1-\lambda) q(\ttheta)] \leq \lambda I[p(\ttheta)] + (1-\lambda) I[q(\ttheta)]
    \end{align}
    for $0 \leq \lambda \leq 1$.
    
    \item $I$ is additive under direct sum, in the sense that if we look at a two probability distributions $p(\ttheta)$ and $q(\ttheta)$, then
    \begin{align}
        I[\lambda p(\ttheta) \oplus (1-\lambda) q(\ttheta)] = \lambda I[p(\ttheta)] + (1-\lambda) I[q(\ttheta)],
    \end{align}
    where the parameter $0 \leq \lambda \leq 1$ ensures that the resulting object is a proper probability distribution.
    
    \item $I$ is non-increasing under stochastic maps. Let $T$ be a stochastic map (a map between probability distributions), then
    \begin{align}
        I[T[p(\ttheta)]] \leq I[p(\ttheta)]
    \end{align}
    which reads as the matrix $I[p(\ttheta)] -I[T[p(\ttheta)]]$ being positive semidefinite.
    
    \item Transformation rule. Suppose we have a reparametrization of the parameters $\ttheta$ given by a multivariate function $\ff(\ttheta)$. The classical Fisher information matrix with respect to the new parameters is given by
    \begin{align}
        I[p(\ff(\ttheta))] = J(\ttheta) I_{\ttheta}[p(\ttheta)] J^{\transp}(\ttheta),
    \end{align}
    where $J_{ij} = \partial \theta_j/\partial f_i$ is the inverse of the Jacobian of the mapping $\ff(\ttheta)$. 
    
    \item Uniqueness as monotone metric. The second order expansion of any monotonic distance measure, \emph{i.e.}\ a distance measure decreasing under stochastic maps, will yield a constant multiple of the classical Fisher information matrix:
    \begin{align}
        d(p(\ttheta), p(\ttheta + \ddelta)) \propto \ddelta^{\transp} I \ddelta + O(\lVert \ddelta \rVert^3).
    \end{align}
\end{enumerate}
\begin{proof}
Recall that the classical Fisher information matrix arises in the second order expansion of the KL divergence,
\begin{align}
    d_{\KL}(p(\ttheta), p(\ttheta + \ddelta)) = \frac{1}{2}\ddelta^{\transp} I \ddelta + O(\lVert \ddelta \rVert^3),
\end{align}
where
\begin{align}\label{eqn:def_cfim_appendix}
    I_{ij} = \left.\frac{\partial^2}{\partial \delta_i \partial \delta_j}d_{\KL}(p(\ttheta), p(\ttheta + \ddelta))\right|_{\ddelta = 0}
\end{align}
\begin{enumerate}[(i)]
\item Symmetry follows from the fact that the derivatives in Eq.~\eqref{eqn:def_cfim_appendix} can be exchanged. Reality is a consequence of the fact that $d_{\KL}$ is a real-valued function.

\item Positive-semidefiniteness follows from the fact that $d_{\KL}$ is a distance measure, meaning that $d_{\KL}(p, q) \geq 0$ and $d_{\KL}(p(\ttheta), p(\ttheta)) = 0$. This implies that we evaluate the second derivative in Eq.~\eqref{eqn:def_cfim_appendix} at a minimum, implying that the curvature in no direction can be negative.

\item Convexity follows from the joint convexity of $d_{\KL}$, \emph{i.e.} the fact that
\begin{align}
    d_{\KL}(\lambda p_1 + (1-\lambda) q_1, \lambda p_2 + (1-\lambda) q_2) \leq \lambda d_{\KL}(p_1, p_2) + (1-\lambda) d_{\KL}(q_1, q_2). 
\end{align}
In our case, we can identify $p_1 = p(\ttheta)$, $q_1 = q(\ttheta)$, $p_2 = p(\ttheta + \ddelta)$ and $q_2 = q(\ttheta + \ddelta)$ and have
\begin{align}
    &d_{\KL}(\lambda p(\ttheta) + (1-\lambda) q(\ttheta), \lambda p(\ttheta + \ddelta) + (1-\lambda) q(\ttheta + \ddelta)) \nonumber \\
    &\qquad \qquad \qquad \leq \lambda d_{\KL}(p(\ttheta), p(\ttheta + \ddelta)) + (1-\lambda) d_{\KL}(q(\ttheta), q(\ttheta + \ddelta)).
\end{align}
The convexity of the classical Fisher information matrix immediately follows by linearity when taking the second derivatives with respect to the components of $\ddelta$ evaluated at $\ddelta = 0$.

\item Due to the direct sum structure, the terms relating to $p(\ttheta)$ and $q(\ttheta)$ in Eq.~\eqref{eqn:formula_cfim_appendix} are independent of each other, yielding the desired result.

\item Monotonicity of the KL divergence implies that under the action of a stochastic map $T$
    \begin{align}
        d_{\KL}(T [p(\ttheta)], T [p(\ttheta + \ddelta)]) \leq d_{\KL}(p(\ttheta), p(\ttheta + \ddelta)),
    \end{align}
    which holds independently of $\ddelta$. In the limit $\lVert \ddelta \rVert \to 0$, we have
    \begin{align}
        d_{\KL}(p(\ttheta), p(\ttheta + \ddelta)) \approx \frac{1}{2}\ddelta^{\transp} I[p(\ttheta)] \ddelta,
    \end{align}
    and therefore
    \begin{align} 
        \ddelta^{\transp} I[T[p(\ttheta)]] \ddelta \leq 
        \ddelta^{\transp} I[p(\ttheta)] \ddelta.
    \end{align}
    As this needs to hold for any $\ddelta$, it implies the sought matrix inequality
    \begin{align}
        I[T[p(\ttheta)]]\leq 
        I[p(\ttheta)].
    \end{align}

\item To prove the transformation rule, we make use of Faà di Bruno's formula for the Hessian. If we have a look at a function $g(\ttheta)$ and seek its second derivatives with respect to coordinates $\ff(\ttheta)$, we have that
\begin{align}
    \frac{\partial^2 g(\ttheta)}{\partial f_i \partial f_j } &= 
    \sum_k \frac{\partial g(\ttheta)}{\partial \theta_k} \frac{\partial^2 u_k}{\partial f_i \partial f_j} + \sum_{kl} \frac{\partial^2 g(\ttheta)}{\partial \theta_k \partial \theta_l} \frac{\partial \theta_k}{\partial f_i} \frac{\partial \theta_l}{\partial f_j}.
\end{align}
If $g(\ttheta)$ is a local optimum (as will be the case for us), the gradient terms vanish and we can rewrite the above using the Jacobian of the coordinate transform $J_{ik} = \partial \theta_k/ \partial f_i$ as
\begin{align}
    \frac{\partial^2 g(\ttheta)}{\partial f_i \partial f_j } &= 
    \sum_{kl} \frac{\partial^2 g(\ttheta)}{\partial \theta_k \partial \theta_l} J_{ik} J_{jl}.
\end{align}
If we denote $H^f_{ij} = \partial^2 g(\ttheta)/\partial f_i \partial f_j$ and $H^{\theta}_{ij} = \partial^2 g(\ttheta)/\partial \theta_i \partial \theta_j$ we can deduce the transformation rule
\begin{align}
    H^f = J H^{\theta} J^{\transp}.
\end{align}
The transformation rule of the classical Fisher information matrix follows from applying this general result to the definition of the classical Fisher information matrix as the Hessian of the underlying distance function.

\item For this result see the original work in Ref.~\cite{morozova1991markov}.
\end{enumerate}
\end{proof}

\section{Modified Distance Functions}\label{app:rescaling}
In this section, we show that applying a post-processing function to a distance function will only change the associated information matrix by a scalar prefactor. 

To do so, we will first have a look at the second derivatives of functions of the form $h(g(\ddelta))$:
\begin{align}
    \frac{\partial^2}{\partial \delta_i \partial \delta_j} h(g(\ddelta)) &=
    \frac{\partial}{\partial \delta_j}\frac{\partial h(g(\ddelta))}{\partial \delta_i} \\
    &= \frac{\partial}{\partial \delta_j}\left( f'(g(\ddelta)) \frac{\partial g(\ddelta)}{\partial \delta_i} \right) \\
    &=f'(g(\ddelta))  \frac{\partial^2 g(\ddelta)}{\partial \delta_i\partial \delta_j} + f''(g(\ddelta)) \left(\frac{\partial g(\ddelta)}{\partial \delta_i}\right)\left(\frac{\partial g(\ddelta)}{\partial \delta_j}\right).
\end{align}

In the case of distance functions $g(\ddelta) = d(\ttheta, \ttheta + \ddelta)$, we want to evaluate them at $\ddelta = 0$ where they are extremal. This means that the first derivatives need to vanish at this point. Therefore, at any extremal point $\ddelta^{*}$, we have that
\begin{align}
    \left.\frac{\partial^2}{\partial \delta_i \partial \delta_j} h(g(\ddelta))\right|_{\ddelta = \ddelta^{*}} = f'(g(\ddelta^{*})) \left. \frac{\partial^2 g(\ddelta)}{\partial \delta_i\partial \delta_j} \right|_{\ddelta = \ddelta^{*}}.
\end{align}
This means that the resulting matrix for $h(g(\ddelta))$ differs from the one for $g(\ddelta)$ only by a factor of $f'(g(\ddelta^{*}))$.

The distance measures $d(\ttheta, \ttheta + \ddelta)$ we are treating in the main part of this work fulfill the extremality property at $\ddelta = 0$ where $d(\ttheta, \ttheta) = 0$. We can therefore say something about the information matrix associated with $h(d)$ compared to $d$:
\begin{align}
    [M_{h(d)}]_{ij} = \left.\frac{\partial^2}{\partial \delta_i \partial \delta_j} h(d(\ttheta, \ttheta + \ddelta))\right|_{\ddelta = 0} = f'(0) \left. \frac{\partial^2 d(\ttheta, \ttheta + \ddelta)}{\partial \delta_i\partial \delta_j} \right|_{\ddelta = 0} = f'(0) [M_{d}]_{ij}
\end{align}
From this formula we immediately see that we need to have that $f'(0) > 0$ to make for a sensible transformation of our information matrix.

The reasoning outlined above also extends beyond functions of the distance. For example, consider the fidelity distance we used to introduce the quantum Fisher information matrix: 
\begin{align}
    d_f(\ttheta, \ttheta + \ddelta) = 1 - |\langle \psi(\ttheta) | \psi(\ttheta + \ddelta) \rangle|^2.
\end{align}
We know that the fidelity $f(\ttheta, \ttheta + \ddelta) = |\langle \psi(\ttheta) | \psi(\ttheta + \ddelta) \rangle|^2$ is extremal for $\ddelta = 0$ where it takes the value 1. We could also define the fidelity distance using the square root of the fidelity -- and some authors do that -- to get
\begin{align}
    d_{\sqrt{f}}(\ttheta, \ttheta + \ddelta) = 1 - |\langle \psi(\ttheta) | \psi(\ttheta + \ddelta) \rangle|.
\end{align}
With our knowledge from above and with seeing that $h(x) = \sqrt{x}$, we immediately see that
\begin{align}
    M_{\sqrt{f}} = h'(f(\ttheta, \ttheta)) M_f = \frac{1}{2} M_f.
\end{align}
This shows that the information matrices of the two conventions only differ by a factor of $1/2$.

\section{Derivation of the Quantum Fisher Information}\label{app:deriv_qfim}
We start the derivation of the quantum Fisher information for pure states from the fidelity distance
\begin{align}
    d_f(\ket{\psi(\ttheta)}, \ket{\psi(\ttheta')}) = 1 - f(\ket{\psi(\ttheta)}, \ket{\psi(\ttheta')}) = 1 - |\langle \psi(\ttheta)|\psi(\ttheta')\rangle|^2.
\end{align}
For small displacements, we have
\begin{align}
    \ket{\psi(\ttheta + \ddelta)} = \ket{\psi(\ttheta)} + \sum_i \delta_i \ket{\partial_i \psi(\ttheta)}.
\end{align}
Therefore,
\begin{align}
    f(\ket{\psi(\ttheta)}, \ket{\psi(\ttheta + \ddelta)}) =
    | \langle \psi(\ttheta) | \psi(\ttheta) \rangle + \sum_i \delta_i \langle \psi(\ttheta) | \partial_i \psi(\ttheta) \rangle |^2 =
    | 1 + \sum_i \delta_i \langle \psi(\ttheta) | \partial_i \psi(\ttheta) \rangle |^2.
\end{align}
Rewriting the absolute value yields
\begin{align}\label{eqn:app_qfim_expansion}
    f(\ket{\psi(\ttheta)}, \ket{\psi(\ttheta + \ddelta)}) = 1 + \sum_i \delta_i (\langle \psi(\ttheta) | \partial_i \psi(\ttheta) \rangle + \langle \partial_i \psi(\ttheta) |  \psi(\ttheta) \rangle)
    + \sum_{ij} \delta_i \delta_j \langle\partial_i \psi(\ttheta) | \psi(\ttheta) \rangle \langle \psi(\ttheta) | \partial_j \psi(\ttheta) \rangle.
\end{align}
We are only interested in second order terms for the computation of our metric, so it would appear that only the last term is of interest to us. But there is a hidden second order dependence in the second term. It arises because $\ket{\psi(\ttheta + \ddelta)} = \ket{\psi(\ttheta)} + \sum_i \delta_i \ket{\partial_i \psi(\ttheta)}$ needs to be a pure state, and therefore
\begin{align}
    1 &= f(\ket{\psi(\ttheta + \ddelta)}, \ket{\psi(\ttheta + \ddelta)}) \\
    &=     \langle \psi(\ttheta)| \psi(\ttheta) \rangle 
    + \sum_i \delta_i (\langle \psi(\ttheta) | \partial_i \psi(\ttheta) \rangle + \langle \partial_i \psi(\ttheta) |  \psi(\ttheta) \rangle)
    + \sum_{ij} \delta_i \delta_j \langle\partial_i \psi(\ttheta) | \partial_j \psi(\ttheta) \rangle,
\end{align}
which enforces
\begin{align}
    \sum_i \delta_i (\langle \psi(\ttheta) | \partial_i \psi(\ttheta) \rangle + \langle \partial_i \psi(\ttheta) |  \psi(\ttheta) \rangle) = -\sum_{ij} \delta_i \delta_j \langle\partial_i \psi(\ttheta) | \partial_j \psi(\ttheta) \rangle.
\end{align}
With these results at hand, we see that
\begin{align}
[M_{f}]_{ij} &=\frac{\partial^2}{\partial \delta_i \partial \delta_j}d_f(\ket{\psi(\ttheta)}, \ket{\psi(\ttheta + \ddelta)}) \\ \label{eqn:hessian_fidelity_distance}
&= 2\operatorname{Re}[\langle\partial_i \psi(\ttheta) | \partial_j \psi(\ttheta) \rangle - \langle\partial_i \psi(\ttheta) | \psi(\ttheta) \rangle \langle \psi(\ttheta) | \partial_j \psi(\ttheta) \rangle].
\end{align}
The real part and the factor 2 account for the fact that $\delta_i \delta_j$ appears in the sum of Eq.~\eqref{eqn:app_qfim_expansion} twice, but with conjugated terms that follow it -- so that we can use that for any complex number $z$
\begin{align}
    z + z^{*} = 2 \operatorname{Re}[z].
\end{align}

Now, we are left with a final step -- we have to ensure that our information matrix is consistent with the classical case. This means that for a \enquote{classical} state with classical measurements, we should recover the classical Fisher information. To this end, let us consider the state 
\begin{align}
    \ket{\psi(\ttheta)} = \sum_l \sqrt{p_l(\ttheta)} \ket{l},
\end{align}
where the sum is over the computational basis states $\ket{l}$. Together with measurements in the computational basis, this state represents the classical probability distribution $p(\ttheta)$. If we put this into the formula we just derived, we see that the first term already resembles the classical Fisher information:
\begin{align}
    \langle\partial_i \psi(\ttheta) | \partial_j \psi(\ttheta) \rangle &= \sum_l (\partial_i \sqrt{p_l(\ttheta)})(\partial_j \sqrt{p_l(\ttheta)}) \\
    &= \frac{1}{4}\sum_l \frac{\partial_i p_l(\ttheta)}{\sqrt{p_l(\ttheta)}}\frac{\partial_j p_l(\ttheta)}{\sqrt{p_l(\ttheta)}} \\
    &= \frac{1}{4}\sum_l\frac{(\partial_i p_l(\ttheta))(\partial_j p_l(\ttheta))}{p_l(\ttheta)}.
\end{align}
And luckily for us, the second term won't contribute, because
\begin{align}
    \langle\partial_i \psi(\ttheta) | \psi(\ttheta) \rangle &= \sum_l  (\partial_i \sqrt{p_l(\ttheta)})\sqrt{p_l(\ttheta)} \\
    &= \frac{1}{2} \sum_l \frac{\partial_i p_l(\ttheta)}{\sqrt{p_l(\ttheta)}}\sqrt{p_l(\ttheta)} \\
    &= \frac{1}{2}\sum_l \partial_i p_l(\ttheta) \\
    &= \frac{1}{2} \partial_i \sum_l p_l(\ttheta) \\
    &= \frac{1}{2} \partial_i 1 \\
    &= 0.
\end{align}
Combining this with Eq.~\eqref{eqn:hessian_fidelity_distance}, we see that -- for the classical state we currently consider -- we have
\begin{align}
    M_{f} &= \frac{1}{2} I.
\end{align}
If we correct for the factor $\frac{1}{2}$ we arrive at the definition of the quantum Fisher information, which is now consistent with the classical limit:
\begin{align}
    \calF_{ij} = 2 [M_{f}]_{ij} = 4 \operatorname{Re} [\langle\partial_i \psi(\ttheta) | \partial_j \psi(\ttheta) \rangle - \langle\partial_i \psi(\ttheta) | \psi(\ttheta) \rangle \langle \psi(\ttheta) | \partial_j \psi(\ttheta) \rangle].
\end{align}

\section{Properties of the Quantum Fisher Information Matrix}\label{app:properties_qfim}
The quantum Fisher Information Matrix $\calF$ of a quantum state $\rho(\ttheta) = \sum_i \lambda_i | \lambda_i \rangle \! \langle \lambda_i |$ with respect to a set of $d$ parameters $\ttheta$ is given by~\cite{liu2020quantum}
\begin{align}\label{eqn:formula_qfim_appendix}
    \calF_{ij} &= \sum_{\substack{kl\\ \lambda_k + \lambda_l \neq 0}} \frac{2 \operatorname{Re}(\langle \lambda_k | \partial_i \rho | \lambda_l \rangle \! \langle \lambda_l | \partial_j \rho | \lambda_k \rangle)}{\lambda_k + \lambda_l} \\
    &= \sum_{\substack{k \\ \lambda_k \neq 0}} \frac{(\partial_i \lambda_k )(\partial_j \lambda_k)}{\lambda_k} + 4 \lambda_k \operatorname{Re}(\langle \partial_i \lambda_k | \partial_j \lambda_k \rangle) - \sum_{\substack{kl \\ \lambda_k, \lambda_l \neq 0}} \frac{8 \lambda_k \lambda_l}{\lambda_k + \lambda_l} \operatorname{Re}(\langle \partial_i \lambda_k | \lambda_l \rangle \! \langle \lambda_l | \partial_j \lambda_k \rangle)
\end{align}
In the case of of a pure state $\rho(\ttheta) = |\psi(\ttheta) \rangle\!\langle \psi(\ttheta) |$ this simplifies to
\begin{align}
    \calF_{ij} = 4 \operatorname{Re}[\langle \partial_i \psi | \partial_j \psi \rangle - \langle \partial_i \psi | \psi \rangle \! \langle \psi | \partial_j \psi \rangle].
\end{align}

The properties of the quantum Fisher information matrix are listed in the excellent review by Liu \emph{et al.}~\cite{liu2020quantum}, but will be given here again for the sake of completeness, along with proofs to foster understanding:
\begin{enumerate}[(i)]
    \item $\calF$ is a real symmetric $d\times d$ matrix.
    \begin{align}
        \calF \in \bbR^{d\times d} \qquad \calF = \calF^{\transp}.
    \end{align}
    
    \item $\calF$ is positive semidefinite, \emph{i.e.}\ it only has non-negative eigenvalues. We write this as
    \begin{align}
        \calF \geq 0.
    \end{align}
    
    \item Convexity. $\calF$ is convex, which means that for any two quantum states $\rho(\ttheta)$ and $\sigma(\ttheta)$,
    \begin{align}
        \calF[\lambda \rho(\ttheta) + (1-\lambda) \sigma(\ttheta)] \leq \lambda \calF[\rho(\ttheta)] + (1-\lambda) \calF[\sigma(\ttheta)]
    \end{align}
    for $0 \leq \lambda \leq 1$.
    
    \item Invariance under unitary transformations. For any unitary transformation $U$, we have that
    \begin{align}
        \calF[U \rho(\ttheta) U^{\dagger}] = \calF[\rho(\ttheta)].
    \end{align}
    
    \item $\calF$ is additive under direct sum, in the sense that if we look at a two quantum states $\rho(\ttheta)$ and $\sigma(\ttheta)$, then
    \begin{align}
        \calF[\lambda \rho(\ttheta) \oplus (1-\lambda) \sigma(\ttheta)] = \lambda \calF[\rho(\ttheta)] + (1-\lambda) \calF[\sigma(\ttheta)],
    \end{align}
    where the parameter $0 \leq \lambda \leq 1$ ensures that the resulting object is a proper quantum state.
    
    \item $\calF$ is additive under tensor products, in the sense that if we look at a two quantum states $\rho(\ttheta)$ and $\sigma(\ttheta)$, then
    \begin{align}
        \calF[\rho(\ttheta) \otimes \sigma(\ttheta)] = \calF[\rho(\ttheta)] + \calF[\sigma(\ttheta)].
    \end{align}
    
    \item $\calF$ is non-increasing under quantum channels. Let $\Phi$ be a quantum channel (a map between density matrices), then
    \begin{align}
        \calF[\Phi[\rho(\ttheta)]] \leq \calF[\rho(\ttheta)]
    \end{align}
    which reads as the matrix $\calF[\rho(\ttheta)] - \calF[\Phi[\rho(\ttheta)]]$ being positive semidefinite.
    
    \item Transformation rule. Suppose we have a reparametrization of the parameters $\ttheta$ given by a multivariate function $\ff(\ttheta)$. The quantum Fisher information matrix with respect to the new parameters is given by
    \begin{align}
        \calF[\rho(\ff(\ttheta))] = J(\ttheta) \calF_{\ttheta}[\rho(\ttheta)] J^{\transp}(\ttheta),
    \end{align}
    where $J_{ij} = \partial \theta_j/\partial f_i$ is the inverse of the Jacobian of the mapping $\ff(\ttheta)$. 
\end{enumerate}
\begin{proof}
In the mixed state case, the quantum Fisher information matrix arises in the second order expansion of the Bures distance. If we have a state $\rho = \sum_i \lambda_i | \lambda_i \rangle \! \langle \lambda_i |$ we denote its unique positive square root as $\rho^{1/2} = \sum_i \sqrt{\lambda_i} | \lambda_i \rangle \! \langle \lambda_i |$. The Bures distance is then given by
\begin{align}
    d_B(\rho, \sigma) = 2 - 2 \Tr \{ (\rho^{1/2} \sigma \rho^{1/2})^{1/2} \}^2
\end{align}
as
\begin{align}
    d_{B}(\rho(\ttheta), \rho(\ttheta + \ddelta)) = \frac{1}{2}\ddelta^{\transp} \calF \ddelta + O(\lVert \ddelta \rVert^3),
\end{align}
where
\begin{align}\label{eqn:def_qfim_appendix}
    \calF_{ij} = \left.\frac{\partial^2}{\partial \delta_i \partial \delta_j}d_{B}(\rho(\ttheta), \rho(\ttheta + \ddelta))\right|_{\ddelta = 0}.
\end{align}

\begin{enumerate}[(i)]
\item Symmetry follows from the fact that the derivatives in Eq.~\eqref{eqn:def_qfim_appendix} can be exchanged. Reality is a consequence of the fact that $d_{B}$ is a real-valued function.

\item Positive-semidefiniteness follows from the fact that $d_{B}$ is a distance measure, meaning that $d_{B}(\rho, \sigma) \geq 0$ and $d_{B}(\rho(\ttheta), \rho(\ttheta)) = 0$. This implies that we evaluate the second derivative in Eq.~\eqref{eqn:def_qfim_appendix} at a minimum, implying that the curvature in no direction can be negative.

\item Convexity follows from the joint convexity of $d_{B}$, \emph{i.e.} the fact that~\cite{fanchini2017geometric}
\begin{align}
    d_{B}(\lambda \rho_1 + (1-\lambda) \sigma_1, \lambda \rho_2 + (1-\lambda) \sigma_2) \leq \lambda d_{B}(\rho_1, \rho_2) + (1-\lambda) d_{B}(\sigma_1, \sigma_2). 
\end{align}
In our case, we can identify $\rho_1 = \rho(\ttheta)$, $\sigma_1 = \sigma(\ttheta)$, $\rho_2 = \rho(\ttheta + \ddelta)$ and $\sigma_2 = \sigma(\ttheta + \ddelta)$ and have
\begin{align}
    &d_{B}(\lambda \rho(\ttheta) + (1-\lambda) \sigma(\ttheta), \lambda \rho(\ttheta + \ddelta) + (1-\lambda) \sigma(\ttheta + \ddelta)) \nonumber \\
    &\qquad \qquad \qquad \leq \lambda d_{B}(\rho(\ttheta), \rho(\ttheta + \ddelta)) + (1-\lambda) d_{B}(\sigma(\ttheta), \sigma(\ttheta + \ddelta)).
\end{align}
The convexity of the quantum Fisher information matrix immediately follows by linearity when taking the second derivatives with respect to the components of $\ddelta$ evaluated at $\ddelta = 0$.

\item The unitary invariance directly follows from the unitary invariance of the Bures distance. To see that the Bures distance is indeed unitarily invariant we first note that
\begin{align}
    (U \rho U^{\dagger})^{1/2} &= \left(\sum_i \lambda_i U | \lambda_i \rangle \! \langle \lambda_i | U^{\dagger} \right)^{1/2} \\
    &=\sum_i \sqrt{\lambda_i} U | \lambda_i \rangle \! \langle \lambda_i | U^{\dagger} \\
    &= U \left(\sum_i \sqrt{\lambda_i} | \lambda_i \rangle \! \langle \lambda_i | \right) U^{\dagger} \\
    &= U \rho^{1/2} U^{\dagger}.
\end{align}
We thus have
\begin{align}
    d_B(U\rho U^{\dagger}, U \sigma U^{\dagger}) &= 2 - 2 \Tr\{ ((U\rho U^{\dagger})^{1/2} U \sigma U^{\dagger} (U\rho U^{\dagger})^{1/2})^{1/2} \}^2 \\
    &=2 - 2 \Tr\{ (U\rho^{1/2} U^{\dagger} U \sigma U^{\dagger} U \rho^{1/2} U^{\dagger})^{1/2} \}^2\\
    &=2 - 2 \Tr\{ U(\rho^{1/2} \sigma  \rho^{1/2} )^{1/2}U^{\dagger} \}^2 \\
    &=2 - 2 \Tr\{ (\rho^{1/2} \sigma  \rho^{1/2} )^{1/2}U^{\dagger}U \}^2 \\
    &=2 - 2 \Tr\{ (\rho^{1/2} \sigma  \rho^{1/2} )^{1/2} \}^2 \\
    &= d_B(\rho,\sigma ).
\end{align}
The unitary invariance of the quantum Fisher information matrix then follows directly from its definition:
\begin{align}
    \calF[U \rho(\ttheta) U^{\dagger}]_{ij} &= \left.\frac{\partial^2}{\partial \delta_i \partial \delta_j}d_{B}(U \rho(\ttheta) U^{\dagger}, U\rho(\ttheta + \ddelta) U^{\dagger})\right|_{\ddelta = 0}\\
    &= \left.\frac{\partial^2}{\partial \delta_i \partial \delta_j}d_{B}(\rho(\ttheta), \rho(\ttheta + \ddelta))\right|_{\ddelta = 0} \\
    &= \calF[\rho(\ttheta)]_{ij}.
\end{align}

\item Due to the direct sum structure, the terms relating to $\rho(\ttheta)$ and $\sigma(\ttheta)$ in Eq.~\eqref{eqn:formula_qfim_appendix} fall into different parts of the sum, yielding the desired result.

\item The tensorization property is a consequence of the tensorization of the Bures fidelity $f_B(\rho, \sigma) = \Tr\{(\rho^{1/2} \sigma  \rho^{1/2} )^{1/2}\}^2$. Note that the matrix square root of a tensor product is the tensor product of the matrix square roots:
\begin{align}
    (\rho \otimes \sigma)^{1/2}=
    (\rho^{1/2} \otimes \sigma^{1/2}).
\end{align}
This implies that the Bures fidelity is multiplicative with respect to tensor products:
\begin{align}
f_B(\rho_1 \otimes \rho_2, \sigma_1 \otimes \sigma_2) &= \Tr\{([\rho_1 \otimes \rho_2]^{1/2} [\sigma_1 \otimes \sigma_2]  [\rho_1 \otimes \rho_2]^{1/2} )^{1/2}\}^2 \\
&= \Tr\{(\rho_1^{1/2} \sigma_1  \rho_1^{1/2} )^{1/2} \otimes (\rho_2^{1/2} \sigma_2  \rho_2^{1/2} )^{1/2}\}^2 \\
&= \Tr\{(\rho_1^{1/2} \sigma_1  \rho_1^{1/2} )^{1/2}\}^2 \Tr \{ (\rho_2^{1/2} \sigma_2  \rho_2^{1/2} )^{1/2}\}^2 \\
&= f_B(\rho_1, \sigma_1)f_B(\rho_2, \sigma_2).
\end{align}
The quantum Fisher information is proportional to the matrix of second derivatives of the Bures fidelity under small perturbation. We can use this to prove the tensorization property:
\begin{align}
    \calF_{ij}[\rho(\ttheta) \otimes \sigma(\ttheta)] &= -2 \left.\frac{\partial^2}{\partial \delta_i \partial \delta_j}f_B(\rho(\ttheta) \otimes \sigma(\ttheta), \rho(\ttheta + \ddelta) \otimes \sigma(\ttheta + \ddelta))\right|_{\ddelta = 0} \\
    &= -2\left.\frac{\partial^2}{\partial \delta_i \partial \delta_j}f_B(\rho(\ttheta), \rho(\ttheta + \ddelta))f_B(\sigma(\ttheta), \sigma(\ttheta + \ddelta))\right|_{\ddelta = 0} \\
    &=-2\Bigg[ \frac{\partial^2 f_B(\rho(\ttheta), \rho(\ttheta + \ddelta))}{\partial \delta_i \partial \delta_j}f_B(\sigma(\ttheta), \sigma(\ttheta + \ddelta)) + f_B(\rho(\ttheta), \rho(\ttheta + \ddelta))\frac{\partial^2 f_B(\sigma(\ttheta), \sigma(\ttheta + \ddelta))}{\partial \delta_i \partial \delta_j} \\
    &+ \frac{\partial f_B(\rho(\ttheta), \rho(\ttheta + \ddelta))}{\partial \delta_i}\frac{\partial f_B(\sigma(\ttheta), \sigma(\ttheta + \ddelta))}{\partial \delta_j}+ \frac{\partial f_B(\rho(\ttheta), \rho(\ttheta + \ddelta))}{\partial \delta_j}\frac{\partial f_B(\sigma(\ttheta), \sigma(\ttheta + \ddelta))}{\partial \delta_i}\Bigg]_{\ddelta = 0}.
\end{align}
We can now exploit the fact that we expand around a minimum, which means that first derivatives vanish. Furthermore, $f_B(\rho, \rho) = 1$ for any $\rho$, which leaves us with only two terms that we can identify with the quantum Fisher information associated to the individual states:
\begin{align}
    \calF_{ij}[\rho(\ttheta) \otimes \sigma(\ttheta)] &= -2 \left[\frac{\partial^2 f_B(\rho(\ttheta), \rho(\ttheta + \ddelta))}{\partial \delta_i \partial \delta_j} + \frac{\partial^2 f_B(\sigma(\ttheta), \sigma(\ttheta + \ddelta))}{\partial \delta_i \partial \delta_j}\right]_{\ddelta = 0} \\
    &= \calF_{ij}[\rho(\ttheta)] + \calF_{ij}[\sigma(\ttheta)],
\end{align}
which concludes the proof.

\item Monotonicity of the Bures divergence implies that under the action of a quantum channel $\Phi$
    \begin{align}
        d_B(\Phi [\rho(\ttheta)], \Phi [\rho(\ttheta + \ddelta)]) \leq d_B(\rho(\ttheta), \rho(\ttheta + \ddelta)),
    \end{align}
    which holds independently of $\ddelta$. In the limit $\lVert \ddelta \rVert \to 0$, we have
    \begin{align}
        d_B(\rho(\ttheta), \rho(\ttheta + \ddelta)) \approx \frac{1}{2}\ddelta^{\transp} \calF[\rho(\ttheta)] \ddelta,
    \end{align}
    and therefore
    \begin{align} 
        \ddelta^{\transp} \calF[\Phi[\rho(\ttheta)]] \ddelta \leq 
        \ddelta^{\transp}\calF[\rho(\ttheta)] \ddelta.
    \end{align}
    As this needs to hold for any $\ddelta$, it implies the sought matrix inequality
    \begin{align}
        \calF[\Phi[\rho(\ttheta)]]\leq 
        \calF[\rho(\ttheta)].
    \end{align}

\item To prove the transformation rule, we make use of Faà di Bruno's formula for the Hessian. If we have a look at a function $g(\ttheta)$ and seek its second derivatives with respect to coordinates $\ff(\ttheta)$, we have that
\begin{align}
    \frac{\partial^2 g(\ttheta)}{\partial f_i \partial f_j } &= 
    \sum_k \frac{\partial g(\ttheta)}{\partial \theta_k} \frac{\partial^2 u_k}{\partial f_i \partial f_j} + \sum_{kl} \frac{\partial^2 g(\ttheta)}{\partial \theta_k \partial \theta_l} \frac{\partial \theta_k}{\partial f_i} \frac{\partial \theta_l}{\partial f_j}.
\end{align}
If the $g(\ttheta)$ is a local optimum (as will be the case for us), the gradient terms vanish and we can rewrite the above using the Jacobian of the coordinate transform $J_{ik} = \partial \theta_k/ \partial f_i$ as
\begin{align}
    \frac{\partial^2 g(\ttheta)}{\partial f_i \partial f_j } &= 
    \sum_{kl} \frac{\partial^2 g(\ttheta)}{\partial \theta_k \partial \theta_l} J_{ik} J_{jl}.
\end{align}
If we denote $H^f_{ij} = \partial^2 g(\ttheta)/\partial f_i \partial f_j$ and $H^{\theta}_{ij} = \partial^2 g(\ttheta)/\partial \theta_i \partial \theta_j$ we can deduce the transformation rule
\begin{align}
    H^f = J H^{\theta} J^{\transp}.
\end{align}
The transformation rule of the quantum Fisher information matrix follows from applying this general result to its definition as the Hessian of the Bures distance.
\end{enumerate}
\end{proof}

\end{document}